\documentclass[prb,twocolumn,notitlepage,superscriptaddress]{revtex4-2}%
\usepackage{amsmath,amsfonts,amssymb,amsthm,epsfig,array}
\usepackage{dsfont}
\usepackage{slashed}
\usepackage{graphics}
\usepackage{float}
\usepackage{verbatim}
\usepackage{color}
\usepackage[dvipsnames]{xcolor}
\usepackage[hidelinks,colorlinks,linkcolor=blue,
citecolor=blue,urlcolor=blue]{hyperref}
\usepackage[titletoc,title]{appendix}
\usepackage{multirow}
\usepackage{bibentry}
\usepackage{tabularx}
\usepackage[mathscr]{euscript}
\usepackage{mathtools}

\usepackage[normalem]{ulem}

\newcommand{\bsl}[1]{\boldsymbol{#1}}

\newcommand{\ii}{\mathrm{i}}
\newcommand{\dd}{\mathrm{d}}

\newcommand{\bra}[1]{\langle #1|}
\newcommand{\ket}[1]{|#1 \rangle}
\newcommand{\braket}[2]{\langle  #1 | #2\rangle}

\newcommand{\dsZ}{\mathbb{Z}}

\newcommand{\dsR}{\mathbb{R}}

\newcommand{\Tr}{\mathop{\mathrm{Tr}}}

\newcommand{\U}[1]{\mathrm{U}(#1)}

\newcommand{\eqnref}[1]{Eq.\,\eqref{#1}}
\newcommand{\figref}[1]{Fig.\,\ref{#1}}
\newcommand{\tabref}[1]{Tab.\,\ref{#1}}

\newcommand{\appref}[1]{Appendix.\,\ref{#1}}
\newcommand{\refcite}[1]{Ref.\,[\onlinecite{#1}]}

\newcommand{\mat}[1]{\left(\begin{matrix}#1\end{matrix}\right)}

\newcommand{\eq}[1]{\begin{equation} #1 \end{equation}}

\newcommand{\eqa}[1]{\begin{align}\begin{split} #1 \end{split}\end{align}}

\let\oldAA\AA
\renewcommand{\AA}{\text{\normalfont\oldAA}}

\newcommand{\ie}{{\emph{i.e.}}}
\newcommand{\eg}{{\emph{e.g.}}}

\newcommand{\Or}[1]{{#1}}
\newcommand{\Bl}[1]{{#1}}

\newcommand{\G}{\mathcal{G}}
\newcommand{\E}{\mathcal{E}}

\newcommand{\s}{\mathrm{s}}
\newcommand{\p}{\mathrm{p}}

\newcommand{\appgendiscu}{Appendix.\,\Bl{A}}

\newcommand{\figtwod}{Fig.\,\Bl{2}}
\newcommand{\eqUfromHgen}{Eq.\,\Bl{(1)} }
\newcommand{\eqDSUmatgen}{Eq.\,\Bl{(4)} }
\newcommand{\eqTwoPlusOneDDynamicalh}{Eq.\,\Bl{(5)} }

\begin{document}
\title{Dynamical Fragile Topology in Floquet Crystals}

\author{Jiabin Yu}
\email{jiabinyu@umd.edu}
\affiliation{Condensed Matter Theory Center and Joint Quantum Institute, Department of Physics, University of Maryland, College Park, MD 20742, USA}

\author{Yang Ge}
\affiliation{Department of Physics, University of Cincinnati, Cincinnati, OH 45221, USA}

\author{Sankar Das Sarma}
\affiliation{Condensed Matter Theory Center and Joint Quantum Institute, Department of Physics, University of Maryland, College Park, MD 20742, USA}

\begin{abstract}
Although fragile topology has been intensely studied in static crystals \Or{in terms of Wannier obstruction}, it is not clear how to generalize the concept to dynamical systems.
In this work, we generalize the concept of fragile topology, and provide a definition of fragile topology for noninteracting Floquet crystals, which we refer to as dynamical fragile topology.
In contrast to the static fragile topology defined by Wannier obstruction, dynamical fragile topology is defined for the nontrivial quantum dynamics characterized by the obstruction to static limits (OTSL).
Specifically, the OTSL of a Floquet crystal is fragile if and only if it disappears after adding a symmetry-preserving static Hamiltonian in a direct-sum way preserving the relevant gaps (RGs).
We further present a concrete 2+1D example for dynamical fragile topology, based on a \Or{model that is qualitatively the same as the dynamical model with anomalous chiral edge modes in [Rudner et al., Phys. Rev. X 3, 031005 (2013)]}.
The fragile OTSL in the 2+1D example exhibits anomalous chiral edge modes for a natural open boundary condition, and does not require any crystalline symmetries besides lattice translations.
Our work paves the way to study fragile topology for general quantum dynamics.

\end{abstract}
\maketitle

\section{Introduction}

In static band insulators, nontrivial topology~\cite{Hasan2010TI,Qi2010TITSC,Chiu2016RMPTopoClas} is defined by Wannier obstruction~\cite{Soluyanov2011WannierZ2,Bradlyn2017TQC,Po2017SymIndi,Kruthoff2017TCI}, \ie, obstruction to the existence of maximally-localized symmetric Wannier functions for the ground state.
\Or{Here maximally-localized symmetric Wannier functions can be intuitively viewed as localized atomic orbitals.
Continuously deforming a topological insulator (by definition having Wannier obstruction) into an atomic insulator (by definition having no Wannier obstruction) must either break certain symmetries or close the gap near the Fermi energy.}
Wannier obstruction of a topological insulator is defined to be fragile~\cite{Po2018FragileTopo,Cano2018DisEBR} if the obstruction disappears after adding an atomic limit to the occupied subspace in a symmetry-preserving way (\figref{fig:gen_dis}(a)).
The K-theoretic classification, as well as the corresponding bulk-boundary correspondence, of stable topology~\cite{Kitaev2009TenFoldWayTITSC,Hasan2010TI,Qi2010TITSC,Freed2013KTheory,Chiu2016RMPTopoClas} fails to fully capture fragile topology, since K-theory~\cite{Hatcher2003VB} requires a stable equivalence which is immune to adding trivial systems.
Therefore, considerable research efforts~\cite{Po2018FragileTopo,Cano2018DisEBR,Wieder2018AXIFragile,Bradlyn2019Fragile,Liu2019ShiftFragile,Slager2019Wilson,Ahn2019TBGFragile,Kooi2019FragileTowfold,Wang2019FragileClassical,dePaz2019FragileLight,Yang2019inversion,Song2020TwistedFragile,Peri2020FragileAcoustic,Song2020FragileAffineMonoid,alexandradinata2020crystallographic,Li2020FragilePhoton,Shang2020FragileCircuitry,Chiu2020FragileFlat,Ji2020FragileAcoustic,Manes2020FragileHoneycomb,Bouhon2020TopologicalMagneticFragile,Bouhon2020FragileGeometric,Wieder2020FragileInducedSemimetal,Skurativska2021FlatBandFragile,Lange2021SubdimensionalFragile} have been dedicated to characterizing and understanding fragile topology during the last three years.
In particular, nontrivial boundary signatures of eigenvalue-indicated fragile phases have been experimentally observed in an acoustic metamaterial with a specially-constructed twisted boundary condition~\cite{Song2020TwistedFragile,Peri2020FragileAcoustic}.
However, it is not straightforward to impose such a twisted boundary condition on naturally occurring condensed matter systems, such as twisted bilayer graphene, which has been predicted to host fragile topology~\cite{Koshino2018TBGFragile,Zou2018TBGFragile,Kang2018TBGFragile,Liu2019TBGFragile,Po2019TBGFragile,Song2019TBGFragile,Ahn2019TBGFragile,Song2020TBGII,Herzog2020HofstadterTopo,Lian2020LLFragileTBG,Peri2021TBGFragileAndSC} (unless the emergent particle-hole symmetry is strictly imposed~\cite{Song2020TBGII}).
Nontrivial~\footnote{Here, for boundary modes to be nontrivial, (i) their spectrum must be inside the bulk gap, and (ii) they are stable even when the underlying Hilbert space includes all irreducible symmetry representations.} boundary signatures of fragile topology for natural open boundary conditions remain elusive~\cite{alexandradinata2020crystallographic}.
Furthermore, all the examples of fragile topology previously studied require crystalline symmetries in addition to lattice translations.

An important open question is how to generalize the concept of fragile topology from static crystals to dynamical systems.
A classic type of dynamical systems are noninteracting Floquet systems---noninteracting systems with time-periodic Hamiltonians---which is the focus of this work.
In recent years, the topology in Floquet systems has been intensely studied~\cite{Sadler2006QuenchecBEC,Oka2009PhotovoltaicHE, Inoue2010FTI, Kitagawa2010TopoFloquet, Lindner2011FTI, Jiang2011MajoranaDriveCAQW, Kitagawa2011PhotoQHI,  Dora2012FSHI, Rudner2013AFTI, Thakurathi2013FloquetMajorana, Wang2013FloquetBloch, Cayssol2013FTI, Rechtsman2013PhotoFTI, Zhou2014FloquetHE, Lababidi2014FloquetQH, Nathan2015TopoSingClas, Keyserlingk2016FloquetSPT, Peng2016AFTISound,Else2016FInteractingTP, Zhao2016FloquetSC,Fruchart2016FloquetTenFold, Potirniche2017FloquetTPCA, Maczewsky2017AFTIPhoto,Mukherjee2017AFTIPhoto,Roy2017FloquetTenFold, Yao2017FloquetTenFold,Eckardt2017FloquetReview, Tarnowski2019FloquetCN, Oka2019FloquetReview, Rudner2020FloquetReview,Nakagawa2020WannierRep, Wintersperger2020ColdAtomAFTI}, especially in the presence of crystalline symmetries~\cite{Morimoto2017FTITimeGlide,Xu2018SpaceTime, Franca2018AHOTI, Rodrigues2019FHOTI, Peng2019FHOTI, Ladovrechis2019AFTCI, Seshadri2019FHOTI, Plekhanov2019FHOTI, Nag2019FHOTI, Bomantara2019FHOTI, Chaudhary2019FHOTIPhononInduced, Ghosh2020FHOTI, Hu2020FHOTI, Huang2020FHOTI, Bomantara2020FMCM, Peng2020FHOTIClassSpaceTimeSym, Bomantara2020FHOTSC, Nag2020FHOT, Ghosh2020FHOTSC, Zhang2020FragileFloquet,Zhu2020FHOTI,Zhang2020FHOTI,Chen2020STSym,Zhu2020FloquetMechanical,Yu2021DSI,Zhu2021SymmetryAFTP}.
In particular, \refcite{Zhang2020FragileFloquet} discussed the fragility of Wannier obstruction in Floquet crystals.
However, nontrivial dynamics of Floquet crystals is characterized by OTSL instead of Wannier obstruction~\cite{Rudner2013AFTI,Peng2016AFTISound,Maczewsky2017AFTIPhoto,Mukherjee2017AFTIPhoto,Nathan2015TopoSingClas, Fruchart2016FloquetTenFold, Roy2017FloquetTenFold, Yao2017FloquetTenFold,Wintersperger2020ColdAtomAFTI,Yu2021DSI}.
Only Floquet crystals with OTSL can exhibit phenomena that are forbidden in static crystals, such as anomalous chiral edge modes~\cite{Rudner2013AFTI} in the absence of nonzero Chern numbers~\cite{TKNN}.
Then, the specific question that we will address is whether the concept of fragile topology can be naturally generalized to the dynamical setting of Floquet crystals with OTSL, regardless of static Wannier obstruction.

In this work, we provide a natural definition of fragile topology with respect to OTSL (\figref{fig:gen_dis}(b)), which we refer to as dynamical fragile topology.
Specifically, the OTSL of a topologically nontrivial Floquet crystal is fragile if and only if the OTSL disappears after adding a symmetry-preserving static Hamiltonian in a direct-sum way that preserves RGs.
Here the precisely-defined RGs are topologically relevant quasienergy band gaps, as explained later.
By definition, dynamical fragile topology cannot be fully captured by K-theory as long as static limits (\ie, Floquet crystals with static Hamiltonians) are treated as trivial systems.
To demonstrate the existence of dynamical fragile topology in tight-binding models, we provide a concrete 2+1D example that has no crystalline symmetries besides lattice translations, based on a slight modification of the model in \refcite{Rudner2013AFTI}.
In particular, the 2+1D example has anomalous chiral edge modes for a natural open boundary condition.
Therefore, unlike static fragile topology, dynamical fragile topology does not rely on crystalline symmetries besides lattice translations, and can have nontrivial boundary signatures for natural open boundary conditions.

\section{General Discussion}

A noninteracting Floquet system can be described by a time-periodic single-particle Hamiltonian
$
H(t)=H(t+T)
$
with $T>0$ the time period, and the corresponding time-evolution operator given by the Dyson series reads
\eq{
\label{eq:U_from_H_gen}
U(t) =\mathcal{T}\exp\left[-\ii \int_0^t dt' H(t')\right]
}
satisfying 
$
U(t+T)=U(t)U(T)
$.
In \eqnref{eq:U_from_H_gen}, $\mathcal{T}$ is the time-ordering operator, $\hbar=1$ is chosen, and the initial time is set to zero without loss of generality~\cite{Yu2021DSI}.

Time-reversal symmetry of $H(t)$ can be naturally broken by the dynamics, and crystals in normal phases typically do not have particle-hole or chiral symmetries; thus we can focus on symmetry class A~\cite{Roy2017FloquetTenFold} \Or{in which time-reversal, particle-hole, and chiral symmetries are all absent.}
Nevertheless, $H(t)$ can preserve a time-independent crystalline symmetry group $\G$.
\Or{Examples of $\G$ include space groups in 3+1D and plane groups in 2+1D, which may contain lattice translation symmetries, rotation symmetries, mirror symmetries, and so on.}
The lattice translation symmetries in $\G$ allow us to label the time-independent bases of the underlying single-particle Hilbert space as $\ket{\psi_{\bsl{k},a}}$, where $\bsl{k}$ is a Bloch momentum in the first Brillouin zone (1BZ), and $a$ takes $N$ different values for all other degrees of freedom (\eg, spin and orbital).
By defining $\ket{\psi_{\bsl{k}}}=(...\ket{\psi_{\bsl{k},a}}...)$, $H(t)$ and $U(t)$ in the space spanned by $\ket{\psi_{\bsl{k}}}$ are represented as
\eq{
\label{eq:mat_rep_H_U}
H(t)=\sum_{\bsl{k}}\ket{\psi_{\bsl{k}}} h(\bsl{k},t) \bra{\psi_{\bsl{k}}},\ U(t)=\sum_{\bsl{k}}\ket{\psi_{\bsl{k}}} U(\bsl{k},t) \bra{\psi_{\bsl{k}}}\ .
}
Eigenvalues of $U(\bsl{k},T)$ have the form $e^{-\ii \E_{m,\bsl{k}} T}$ with $m=1,...,N$, and $\E_{m,\bsl{k}}$ are called quasienergy bands.

\begin{figure}
    \centering
    \includegraphics[width=\columnwidth]{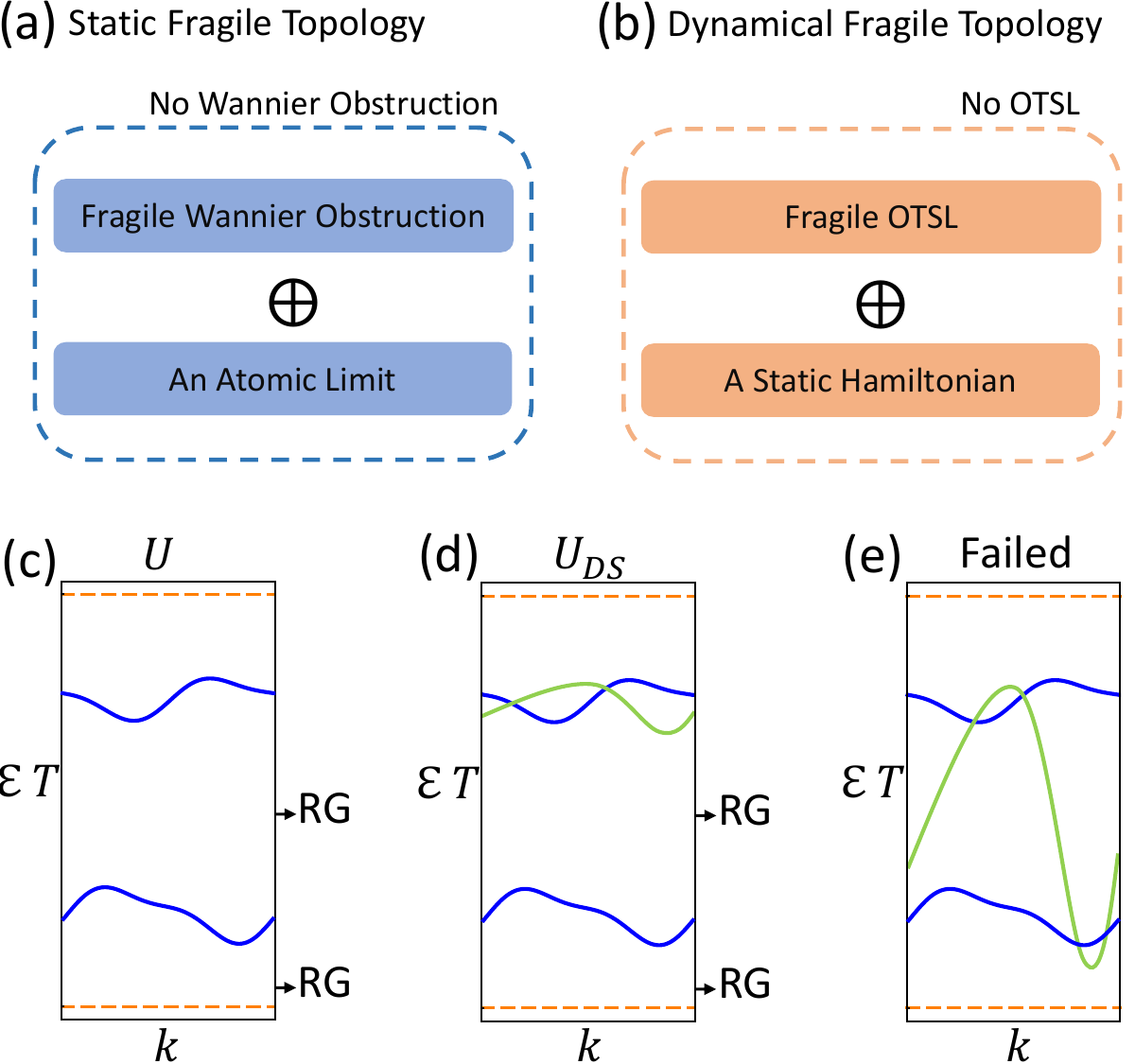}
    \caption{
    In (a-b), we schematically show the definitions of static fragile topology and dynamical fragile topology.
    In (c), we schematically plot the quasienergy bands for a two-band 1+1D Floquet crystal $U$, for which we choose both quasienergy gaps to be RGs.
    In either (d) or (e), we schematically plot the quasienergy bands of a direct-sum system that consists of $U$ (blue) and an added one-band static Hamiltonian (green).
    In (d), both RGs of $U$ are preserved, while one RG of $U$ is closed in (e).
    In (c-e), all quasienergy bands are plotted in a time-independent phase Brillouin zone (PBZ) $[\Phi_{\bsl{k}},\Phi_{\bsl{k}}+2\pi)$ with the PBZ lower bound $\Phi_{\bsl{k}}$ in a RG (if RGs exist).
    The orange dashed lines mark the boundaries of the PBZ.
    }
    \label{fig:gen_dis}
\end{figure}

The quasienergy band gaps play an particularly important role in Floquet topology~\cite{Nathan2015TopoSingClas,Roy2017FloquetTenFold,Yao2017FloquetTenFold,Yu2021DSI}, similar to that of energy band gaps in static band topology.
Nevertheless, unlike a static band insulator whose physically relevant band gap is uniquely determined by the filling, Floquet systems like $U$ do not have a well-defined occupied subspace, and thereby we have to choose relevant quasienergy band gaps (\ie, RGs) for them based on the physics of interest.
In other words, choice of RGs is an essential step in describing Floquet topology.
After choosing RGs for $U$, we arrive at a Floquet crystal $U$ that is characterized by its time-evolution operator $U(t)$ equipped with the time period $T$, the RG choice, and the crystalline symmetry group $\G$. (See \figref{fig:gen_dis}(c) for a schematic example.)

\Or{According to \refcite{Nathan2015TopoSingClas,Yu2021DSI}, the topological equivalence between two $\G$-invariant Floquet crystals is defined by a continuous deformation that connects them while preserving $\G$ and all RGs.
Then, as proposed in \refcite{Yu2021DSI},} a Floquet crystal is defined to have OTSL if and only if it is topologically distinct from all $\G$-invariant static limits.\cite{SM}
\Or{
In other words, given a Floquet crystal $U$ with $\G$, it has OTSL if and only if we cannot continuously deform $U$ into the time-evolution operator of any static Hamiltonian while preserving all symmetries in $\G$ and keeping open all RGs of $U$.
For example, the 2+1D two-band dynamical model in \refcite{Rudner2013AFTI}, which has zero Chern numbers and has anomalous chiral edge modes, has OTSL if all bulk quasienergy gaps are chosen to be RGs.
The reason is that 2+1D static systems cannot have chiral edge modes when all bands have zero Chern numbers, and thus connecting the 2+1D dynamical model to any 2+1D static crystal in a symmetry-preserving way must close certain RGs to change the Chern numbers.
For general 2+1D Floquet systems with $\G$ containing only lattice translation symmetries, \refcite{Rudner2013AFTI,Nathan2015TopoSingClas,Roy2017FloquetTenFold} suggests to classify them by a winding number $W$ defined in \refcite{Rudner2013AFTI} and the Chern numbers of the bulk quasienergy bands.
Based on \refcite{Rudner2013AFTI}, a nonzero $W$ and zero Chern numbers can indicate OTSL when all bulk quasi-energy gaps are relevant, since a nonzero $W$ in this case can indicate the existence of anomalous chiral edge modes. 
But a nonzero $W$ itself does not necessarily indicate OTSL, since static systems can have nonzero $W$'s as \refcite{Rudner2013AFTI} pointed out.

One experimental signature of OTSL (though not conclusive) is the closing of certain RGs when deforming the dynamical system to any static limit while preserving $\G$.
We believe this signature of OTSL is experimentally accessible since tracking the quasienergy spectrum while deforming the systems has been achieved in experiments like \refcite{Wintersperger2020ColdAtomAFTI}. 
}

We now define dynamical fragile topology (\figref{fig:gen_dis}(b)).
Suppose the Floquet crystal $U$ has OTSL.
Its OTSL is defined to be fragile if and only if the OTSL disappears after adding a $\G$-invariant static Hamiltonian $H_{SL}$ in a direct-sum way that preserves all RGs.
\Or{$H_{SL}$ is allowed to have additional symmetries that are absent in the Floquet crystal $U$, but they are irrelevant to our discussion.}
The direct-sum way is required by K-theory, suggesting that the bases of $H_{SL}$, denoted by $\ket{\psi_{\bsl{k}}^{SL}}=(...\ket{\psi_{\bsl{k},a_{SL}}^{SL}} ...)$, must be orthogonal to the bases $\ket{\psi_{\bsl{k},a}}$ of $U$.
Then, the underlying Hilbert space of the direct-sum Hamiltonian $H_{DS}(t)=H(t) + H_{SL}$ is spanned by $\ket{\psi_{\bsl{k}}^{DS}}=(\ket{\psi_{\bsl{k}}},\ket{\psi_{\bsl{k}}^{SL}})$. 
In this direct-sum space, the time-evolution operator $U_{DS}(t)$ is represented as 
\eq{
\label{eq:DS_U_gen}
U_{DS}(t)=\sum_{\bsl{k}} \ket{\psi_{\bsl{k}}^{DS}} U_{DS}(\bsl{k},t) \bra{\psi_{\bsl{k}}^{DS}}\ ,
}
where 
\eq{
\label{eq:DS_U_mat_gen}
U_{DS}(\bsl{k},t)=\mat{U(\bsl{k},t) & \\ & e^{-\ii h_{SL}(\bsl{k}) t} }\ ,
}
and $h_{SL}(\bsl{k})$ is the continuous representation of $H_{SL}$ furnished by $\ket{\psi_{\bsl{k}}^{SL}}$.
Since $H_{SL}$ is $\G$-invariant, $U_{DS}(t)$ also preserves $\G$.

Preserving RGs means that all RGs of $U$ are kept open in the quasienergy band structure given by $U_{DS}(\bsl{k},T)$. (See a RG-preserving example $U_{DS}$ in \figref{fig:gen_dis}(d) and nonpreserving example in \figref{fig:gen_dis}(e).)
Then, we can choose the RGs of $U_{DS}$ to be the same as those of $U$.
Combined with the time period $T$ and the crystalline symmetry group $\G$, we now have a direct-sum Floquet crystal $U_{DS}$.
The absence of OTSL in turn means that $U_{DS}$ has no OTSL, or equivalently $U_{DS}(t)$ can be continuously deformed into the time-evolution operator of a static Hamiltonian without breaking symmetries in $\G$ and without closing any RGs of $U_{DS}$.
Although there is no off-diagonal coupling in \eqnref{eq:DS_U_mat_gen}, symmetry-preserving couplings are allowed when constructing the deformation of $U_{DS}$.
Crucially, as long as we can find one $H_{SL}$ that yields a direct-sum $U_{DS}$ without OTSL, the OTSL of $U$ is fragile.

We emphasize that any fragile OTSL that satisfies the above definition is still fragile even if we allow $U_{DS}$ to close RGs of $U$.
The intuition is that if $U_{DS}$ closes certain RGs of $U$ (like \figref{fig:gen_dis}(e)) and we choose the remaining RGs of $U$ as RGs of $U_{DS}$, it would be easier for $U_{DS}$ to lose OTSL since the deformation of $U_{DS}$ is constrained by fewer RGs.
Furthermore, our general discussion does not rely on specific RG choices for the dynamical $U$.
One straightforward RG choice for the dynamical $U$ is taking all quasienergy gaps to be relevant, which has been adopted in both theoretical~\cite{Nathan2015TopoSingClas} and experimental~\cite{Wintersperger2020ColdAtomAFTI} works, while other choices are also consistent with the above definition.
In general, there are no efficient methods of determining fragile OTSL, because there is no rigorously-proven complete topological classification for generic Floquet crystals with arbitrary crystalline symmetry groups.
In other words, even if the direct-sum $U_{DS}$ has trivial topological invariants according to the currently-known classification, there is no proof that $U_{DS}$ must have no OTSL.
Therefore, we cannot tell from the known classification whether fragile OTSL exists in tight-binding models.
In order to prove the existence, we present below a concrete example demonstrating fragile OTSLs.

\begin{table}[t]
    \centering
    \begin{tabular}{c|ccc}
        $t$ & $[\frac{T}{5}, \frac{2T}{5})$ & $[\frac{2T}{5}, \frac{3T}{5})$  & $[\frac{3T}{5}, \frac{4T}{5})$ \\
        \hline
         $d_x$ &   $-1.25\cos(k_y)$ & $-1.25\cos(k_x-k_y)$ & $-1.25\cos(k_x)$ \\ 
         $d_y$ &   $1.25\sin(k_y)$ & $-1.25\sin(k_x-k_y)$ & $-1.25\sin(k_x)$ \\ 
    \end{tabular}
    \caption{The nonzero expressions of $d_{x,y}$ in the Hamiltonian in \eqnref{eq:h_2D} within one time period.
    $d_{x,y}=0$ for $t\in [0, \frac{T}{5})\cup [\frac{4T}{5}, T)$.}
    \label{tab:2D_h_app}
\end{table}

\section{2+1D Example with p1 Plane Group}

\begin{figure}[t]
    \centering
    \includegraphics[width=\columnwidth]{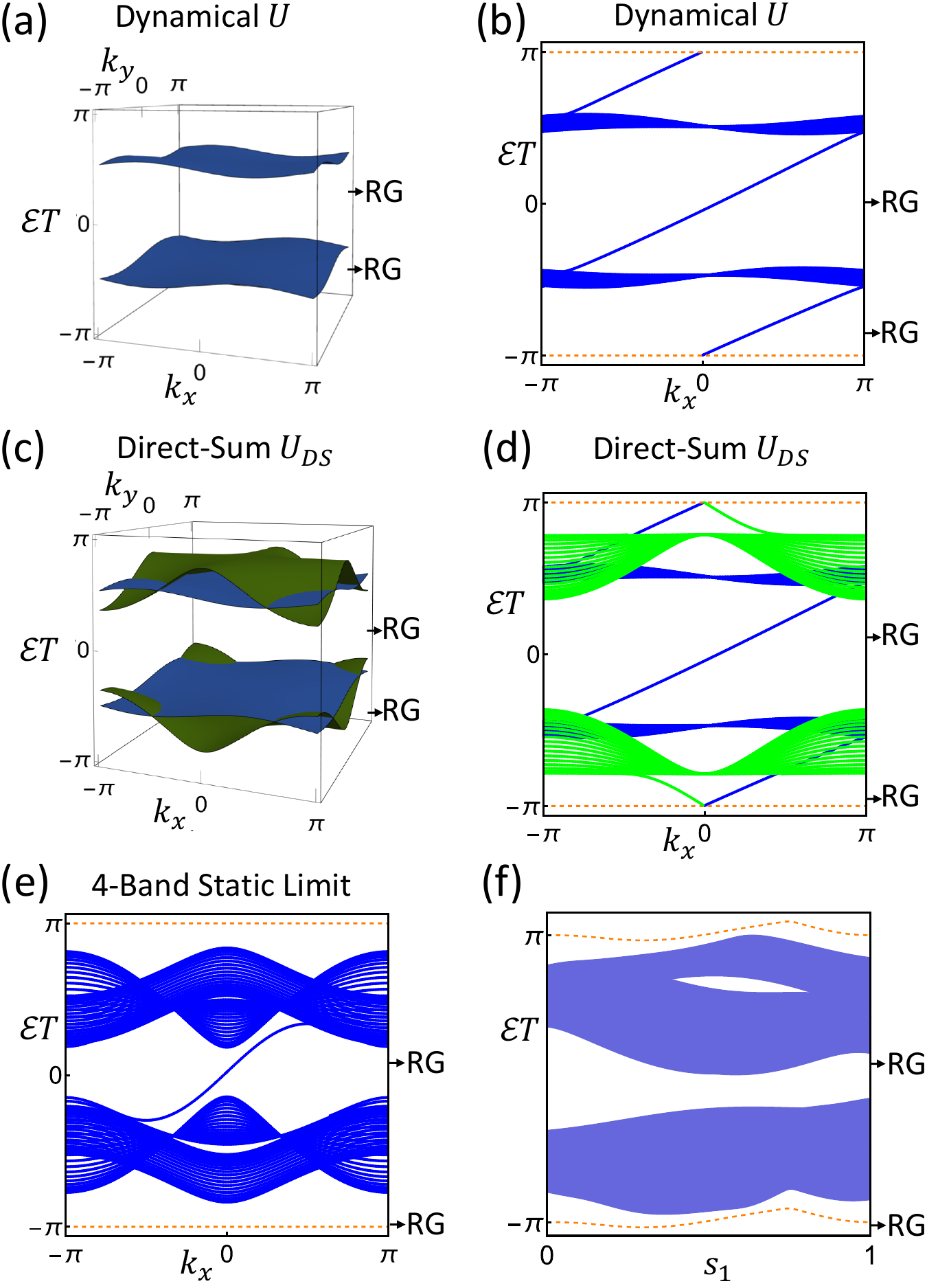}
    \caption{
    In this figure, we show the dynamical fragile topology in the 2+1D example.
    All RGs are labeled according to the bulk quasienergy bands. 
    In (a) and (c), we plot the bulk quasienergy bands of dynamical $U$ and direct-sum $U_{DS}$, respectively, in the PBZ $[-\pi,\pi)$.
    In (b), (d) and (e), we plot quasienergy bands of the dynamical $U$, the direct-sum $U_{DS}$ and a four-band static limit, respectively, for an open boundary condition along $y$.
    Specifically, we choose $20$ lattice sites along $y$, and only include bulk states (dense lines) and the $(0\bar{1})$-edge states (in-gap isolated lines) for (b,d,e).
    The orange dashed lines are the PBZ boundaries.
    In (c-d), the quasienergy bands given by the dynamical $U$ and the added static Hamiltonian $H_{SL}$ are respectively marked in blue and green.
    In (f), we show the quasienergy range (purple regions) of bulk quasienergy bands derived from the deformation $\widetilde{U}_{2D,s_1}(T)$.
    The white regions indicate the deformed RGs, and the orange dashed lines are the deformed PBZ boundaries.
    }
    \label{fig:2D_p1}
\end{figure}

Based on a slightly modified version of the model in \refcite{Rudner2013AFTI}, we introduce a 2+1D example that has fragile OTSL.
In particular, we will demonstrate that dynamical fragile topology can have nontrivial boundary signatures for a natural open boundary condition.
The crystalline symmetry group for this example is $\G= p1$, which only contains lattice translations.
For the dynamical model, we consider a square lattice with two spinless localized orbitals (labeled by $1,2$) on each lattice site, resulting in the bases $\ket{\psi_{\bsl{k}}}=(\ket{\psi_{\bsl{k},1}},\ket{\psi_{\bsl{k},2}})$.
With these bases, the two-band tight-binding Hamiltonian $H(t)$ that we use is represented as 
\eq{
\label{eq:h_2D}
h(\bsl{k},t)=E_0 \sigma_0+ d_x(\bsl{k},t) \sigma_x+ d_y(\bsl{k},t) \sigma_y+ \delta \sigma_z\ ,
}
where $h(\bsl{k},t+T)=h(\bsl{k},t)$ with $T=2\pi$ determining the unit of energy, the lattice constant is set as the unit of length, $E_0=0.01$, $\delta=0.1$, and the detailed expressions of $d_x$ and $d_y$ are shown in \tabref{tab:2D_h_app}.

The time-evolution operator $U(t)$ and time-evolution matrix $U(\bsl{k},t)$ in the space spanned by $\ket{\psi_{\bsl{k}}}$ can be derived from \eqnref{eq:U_from_H_gen}-\eqref{eq:mat_rep_H_U}.
We plot the bulk quasienergy bands of the dynamical $U$ in \figref{fig:2D_p1}(a), showing two bulk quasienergy bands and two bulk quasienergy gaps.
By choosing both bulk quasienergy gaps to be RGs, we complete a Floquet crystal $U$ with time period $T$ and $\G=p1$. 
Direct calculation shows that each bulk quasienergy band has a zero Chern number.
We further plot the quasienergy bands of $U$ for an open boundary condition along $y$ in \figref{fig:2D_p1}(b), which shows one chiral gapless mode in each bulk RG at each edge.
According to \refcite{Rudner2013AFTI}, such chiral edge modes are anomalous, and $U$ must have OTSL, because chiral edge modes are forbidden in static systems with only vanishing Chern numbers~\cite{SM}.
\Or{
We emphasize that the model \eqnref{eq:h_2D} is qualitatively the same as the dynamical two-band model with anomalous edge modes in \refcite{Rudner2013AFTI}, since (i) both models have two bulk quasienergy bands with zero Chern numbers, (ii) both models have one anomalous chiral edge mode at each edge in each bulk quasienergy gap, and (iii) both models have the winding number $W$ defined in \refcite{Rudner2013AFTI} being 1 \cite{SM}.
The explicit differences between them, which are minor,  are detailed in \cite{SM}.
}

Next we show that the OTSL of $U$ is fragile.
To do so, we need to add a static Hamiltonian to form a direct-sum Floquet crystal.
The static Hamiltonian that we add is a two-band static tight-binding Hamiltonian $H_{SL}$ on the same square lattice as the dynamical $U$.
We consider two localized orbitals (labeled as 3,4) on each lattice site for $H_{SL}$, and make sure that the resultant bases $\ket{\psi^{SL}_{\bsl{k}}}=(\ket{\psi_{\bsl{k},3}},\ket{\psi_{\bsl{k},4}})$ are orthogonal to the bases of the dynamical $U$.
We choose the matrix representation of $H_{SL}$ as 
\eq{
\label{eq:2+1D_SL_h}
h_{SL}(\bsl{k})=\frac{1}{2}+\frac{1}{3\pi}[M(\bsl{k})\sigma_z+\sin(k_x)\sigma_x+\sin(k_y)\sigma_y]\ ,
}
where $M(\bsl{k})=\cos(k_x)+\cos(k_y)-1$, and $h_{SL}(\bsl{k})$ represents a simple lattice model for the quantum anomalous Hall effect~\cite{Haldane1988QAH}.
\Or{Quantum anomalous Hall effect has been experimentally realized~\cite{Chang2013QAH,Zhao2020QAHHigherCN}.}
According to \eqnref{eq:DS_U_gen}, the bases of the direct-sum Hamiltonian $H_{DS}(t)=H(t)+H_{SL}$ are $\ket{\psi^{DS}_{\bsl{k}}}=(\ket{\psi_{\bsl{k}}},\ket{\psi_{\bsl{k}}^{SL}})$, and the matrix representation $U_{DS}(\bsl{k},t)$ of the time-evolution operator $U_{DS}(t)$ can be derived by \eqnref{eq:DS_U_mat_gen}.
As shown in \figref{fig:2D_p1}(c), the four bulk quasienergy bands given by $U_{DS}(\bsl{k},T)$ keep open both bulk RGs of $U$.
We can then choose the bulk RGs of $U$ to be the RGs of $U_{DS}$, giving us a direct-sum Floquet crystal $U_{DS}$ together with $T$ and $\G=p1$.

In particular, the bulk bands of the static Hamiltonian $H_{SL}$ have nonzero Chern numbers, which give one chiral edge mode on each edge (\figref{fig:2D_p1}(d)).
The chiral edge mode brought by $H_{SL}$ crosses with a counter-propagating chiral edge mode of the dynamical $U$ in the lower bulk RG.
As the crossing is unstable, the structure of the chiral modes suggests that the direct-sum $U_{DS}$ might be topologically equivalent to a four-band static limit shown in \figref{fig:2D_p1}(e).
To confirm it, we construct a continuous deformation $\widetilde{U}_{2D,s_1}(t)$ from $U_{DS}$ $(s_1=0)$ to the four-band static limit $(s_1=1)$.
As shown in \figref{fig:2D_p1}(f), both RGs of $U_{DS}$ are kept open along the deformation $(s_1\in[0,1])$ and become the RGs of the four-band static limit at $s_1=1$, indicating that $U_{DS}$ is topologically equivalent to the four-band static limit and thus has no OTSL~\cite{SM}.
Therefore, OTSL of the dynamical $U$ is fragile, and the fragile OTSL has anomalous chiral edge modes for a natural open boundary condition.
Moreover, the fragile OTSL does not need any crystalline symmetries besides lattice translations.
    
\section{Physical Implications}

We now discuss the physical implications of the defined dynamical fragile topology.
One way~\cite{Oka2009PhotovoltaicHE,Wang2013FloquetBloch,Peng2020FloquetMajoranaPlanarJosephson,Zhang2020AFCTSC} to generate a well-defined Floquet system is applying a temporal drive to certain modes in a static system, while leaving the other modes (effectively) static.
For example, \refcite{Wang2013FloquetBloch} used laser to excite only the low-energy electrons in a sample, leaving high-energy modes (effectively) undriven.
For this method, it is important to carefully control the applied drive (like carefully choosing the laser wavelength in the above example~\cite{Wang2013FloquetBloch}) so that the driven modes have negligible coupling to the remaining static modes.
Then, the driven modes form an isolated Floquet subsystem, which can have well-defined OTSL.

One natural question is whether the OTSL in such a Floquet subsystem still exists if the coupling between the subsystem and the surrounding static modes becomes strong.
Our definition provides a formalism to address this question.
Specifically, when the dynamical $U$ is a Floquet subsystem with OTSL, the added static Hamiltonian $H_{SL}$ will correspond to the surrounding static modes that might have large coupling to $U$.
In a physical situation where the coupling between $U$ and the surrounding static modes is allowed to be nonzero, the nonzero coupling may introduce nonzero off-diagonal coupling in \eqnref{eq:DS_U_mat_gen}.
Then, the OTSL of the entire system should be determined by the direct-sum $U_{DS}$ rather than $U$, since $U$ is not isolated and the nonzero off-diagonal coupling in \eqnref{eq:DS_U_mat_gen} can be naturally included during the deformation of $U_{DS}$.
In this case, if $U_{DS}$ has no OTSL, it means that the OTSL in $U$ can be destroyed by the surrounding static modes, and thereby is fragile.
In other words, if the Floquet subsystem $U$ has fragile OTSL, its OTSL will disappear when certain surrounding static modes are included.

Even if the dynamical $U$ is not a subsystem of a larger system, it is also possible to test the fragility of OTSL in $U$ by explicitly adding a static Hamiltonian.
For example, the 2+1D Floquet topological phases with anomalous edge modes have been observed in a cold-atom system~\cite{Wintersperger2020ColdAtomAFTI} and a photonic system~\cite{Maczewsky2017AFTIPhoto}.
In the photonic waveguide arrays, the fragility of OTSL may be tested by adding a set of straight waveguides as a static Hamiltonian~\cite{Rechtsman2013PhotoFTI,Ozawa2019TopoPhotonicsRMP}.

\section{Conclusion and Discussion}

To sum up, we introduce a definition for dynamical fragile topology with respect to OTSL, and present a concrete 2+1D example.
The 2+1D example shows that dynamical fragile topology does not rely on crystalline symmetries other than just lattice translations, and can have clear boundary signatures---such as anomalous chiral edge modes---for a natural open boundary condition.

In the presence of crystalline symmetries beyond lattice translations, dynamical fragile topology can also exist in tight-binding models.
To demonstrate this point, we construct a 1+1D model with inversion symmetry, which is carefully discussed in \refcite{SM}.
Briefly speaking, the 1+1D dynamical model with OTSL is given by a chain of spinless s and p orbitals with time-dependent onsite energy and nearest-neighbor hopping, and its OTSL disappears after adding a static chain of spinless d orbitals.
\Or{This example also shows that the fragile OTSL in certain cases can be destroyed by adding a static atomic insulator.
We emphasize that although we destroyed the fragile OTSL in the 2+1D example by adding a Chern insulator, we cannot rule out the possibility that adding certain static atomic insulators may also do the job.
Finding such static atomic insulators for the 2+1D example would be an interesting future direction.
}
Furthermore, although our work focuses on symmetry class A, the definition of dynamical fragile topology can be generalized to other symmetry classes by including more internal symmetries.

Finally, we compare and contrast our results to \refcite{Roy2017FloquetTenFold}.
\refcite{Roy2017FloquetTenFold} presented a K-theoretic classification of unitary loops (\ie, time-periodic unitary evolution) of Floquet crystals.
We emphasize that \refcite{Roy2017FloquetTenFold} defined dynamical topological systems by nontrivial unitary loops, while we use the OTSL definition proposed in \refcite{Yu2021DSI}.
Having nontrivial unitary loops is not equivalent to having OTSL, because static limits may have nontrivial unitary loops. (See \appgendiscu\ of this work and Appendix C of \refcite{Rudner2013AFTI}.)
Due to the different definitions, the dynamical model in the 2+1D example is identified as stable dynamical topological by the K-theoretic classification in \refcite{Roy2017FloquetTenFold}, while we find a fragile OTSL in it.~\cite{SM}

\section{Acknowledgement}

J.\,Yu thanks Yu-An Chen, Biao Lian, Zhi-Da Song, Xiao-Qi Sun, Zhi-Cheng Yang and Rui-Xing Zhang for helpful discussions. 
In particular, J.\,Yu thanks Zhi-Da Song for providing critical comments, and thanks Xiao-Qi Sun for the information on straight waveguides as static Hamiltonians in photonic systems.
This work is supported by the Laboratory for Physical Sciences.

\bibliography{bibfile_references}

\clearpage

\begin{widetext}

\tableofcontents

\appendix

\section{Details on Topological Equivalence and OTSL}
\label{app:gen_discu}

In this section, we following \refcite{Yu2021DSI} to review the definitions of topological equivalence and OTSL for Floquet crystals in the symmetry class A.
We also present a static limit with nontrivial unitary loop at the end.

Within this section, we use $s\in[0,1]$ to label the tuning parameter for the continuous deformation, which should not be confused with the $\s$ orbital mentioned in the main text.
Moreover, we always imply that the Bloch momentum $\bsl{k}$ takes values in 1BZ, unless explicitly stated otherwise.

Before reviewing the two definitions and discussing the example, let us first elaborate on the requirements for time-independent bases $\ket{\psi_{\bsl{k},a}}$ of a Hilbert space $\mathcal{H}$ of a $d+1D$ Floquet crystal, which we have implied and will still imply throughout the work.
First, the bases must be orthonormal: $\braket{\psi_{\bsl{k}',a'}}{\psi_{\bsl{k},a}}=\delta_{\bsl{k}'\bsl{k}}\delta_{a'a}$.
Second, the bases must be complete in $\mathcal{H}$: $\sum_{\bsl{k},a}\ket{\psi_{\bsl{k},a}}\bra{\psi_{\bsl{k},a}}=1$ in $\mathcal{H}$.
Third, $\ket{\psi_{\bsl{k}+\bsl{K},a}}=\ket{\psi_{\bsl{k},a}}$ for all reciprocal lattice vectors $\bsl{K}$.
Fourth, the periodic parts of chosen bases $e^{-\ii \bsl{k}\cdot\bsl{r}}\ket{\psi_{\bsl{k},a}}$ are smooth functions of $\bsl{k}\in\dsR^d$.
The third convention allows us to extend the domain of $\bsl{k}$ from 1BZ to $\dsR^d$ for any matrix representation furnished by $\ket{\psi_{\bsl{k},a}}$, through making the representation invariant under $\bsl{k}\rightarrow \bsl{k}+\bsl{K}$, which we have implied throughout this work.
We also imply that $a$ in $\ket{\psi_{\bsl{k},a}}$ only takes a finite number of different values.

In particular, given a Hilbert space $\mathcal{H}$ spanned by real-space local bases $\ket{\bsl{R},a}$ of any tight-binding model, 
\eq{
\ket{\psi_{\bsl{k},a}}=\frac{1}{\sqrt{\mathcal{N}}}\sum_{\bsl{R}}e^{\ii \bsl{k}\cdot\bsl{R}}\ket{\bsl{R},a}
} 
are always bases of $\mathcal{H}$ that satisfy the above four requirements, where $\mathcal{N}$ is the number of lattice sites.
The above expression has been used to derive the bases for all the examples in this work.
Furthermore, given $\ket{\psi_{\bsl{k}}}=(...\ket{\psi_{\bsl{k},a}}...)$ as bases for $\mathcal{H}$ and $\ket{\psi_{\bsl{k}}'}=(...\ket{\psi_{\bsl{k},a'}'}...)$ as bases for $\mathcal{H}'$ and $\braket{\psi_{\bsl{k}',a'}'}{\psi_{\bsl{k},a}}=0$, then $\ket{\psi_{\bsl{k}}^{DS}}=(\ket{\psi_{\bsl{k}}},\ket{\psi_{\bsl{k}}'})$ are bases of the space spanned by $\ket{\psi_{\bsl{k}}^{DS}}$, which satisfy the above four requirements.

\subsection{Topological Equivalence}

Given two $d+1$D Floquet crystals that are invariant under the same crystalline symmetry group $\G$, denoted by $U$ (with $T$, a RG choice, and $\G$) and $U'$ (with $T'$, a RG choice, and $\G$).
$U$ and $U'$ are topologically equivalent if and only if there exits a continuous deformation $U_s(t)$ with $s\in [0,1]$ such that $U_s(t)$ is unitary, $U_{s=0}(t)=U(t)$, $U_{s=1}(t)=U'(t)$, and $U_s(t)$ preserves $\G$ and all RGs.
We will elaborate on the meaning of being continuous, preserving $\G$, and preserving RGs in the following. 

Being continuous means that (i) $U_s(t+T_s)=U_s(t)U_s(T_s)$, where $T_s>0$ is a continuous function of $s$, $T_{s=0}=T$, and $T_{s=1}=T'$, and (ii) there exist $\ket{\psi_{s,\bsl{k}}}=(...,\ket{\psi_{s,\bsl{k},a}},...)$ as bases of $U_s(t)$ at each value of $s$ such that $e^{-\ii \bsl{k}\cdot\bsl{r}}\ket{\psi_{s,\bsl{k}}}$ is a continuous function of $(\bsl{k},s)\in \dsR^d\times [0,1]$, and the matrix representation of $U_s(t)$ labeled as $U_s(\bsl{k},t)$ is a continuous function of $(\bsl{k},t,s)\in \dsR^d\times\dsR\times [0,1]$.
Note that $\ket{\psi_{s,\bsl{k},a}}$ as bases must be orthonormal for $(\bsl{k},a)$, must be complete in the Hilbert space of $U_s(t)$ at each $s$, must have periodic part $e^{-\ii \bsl{k}\cdot\bsl{r}}\ket{\psi_{s,\bsl{k},a}}$ smooth for $\bsl{k}\in\dsR^d$, and must satisfy $\ket{\psi_{s,\bsl{k}+\bsl{K},a}}=\ket{\psi_{s,\bsl{k},a}}$ for all reciprocal lattice vectors $\bsl{K}$.
Preserving $\G$ means that for all $g\in\G$, $[U_s(t),g]=0$ and $g\ket{\psi_{s,\bsl{k}}}=\ket{\psi_{s,\bsl{k}_g}} u_{s,g}(\bsl{k})$, where $\bsl{k}_g=R\bsl{k}$ and $R$ is the point-group part of $g$.
Preserving RGs means that the RGs of $U$ are kept open as tuning $s$ continuously from $s=0$ to $s=1$, and they become the RGs of $U'$ at $s=1$.

In the two examples discussed in the main text, we use one special type of topological equivalence, which is discussed in the following.
Given a $d+1$D Floquet Hamiltonian $H(t)$ with crystalline symmetry group $\G$.
Suppose it generates a Floquet crystal $U$ with time period $T$, a RG choice, and $\G$, and the underlying Hilbert space is spanned by the bases $\ket{\psi_{\bsl{k}}}$.
Then, we have $h(\bsl{k},t)$ as the representation of $H(t)$ furnished by $\ket{\psi_{\bsl{k}}}$.
Now suppose we have a Hermitian $H_s(t)$ as
\eq{
H_s(t)=\sum_{\bsl{k}} \ket{\psi_{\bsl{k}}} h_s(\bsl{k},t)\bra{\psi_{\bsl{k}}}
}
with $s\in[0,1]$; $H_s(t)$ satisfies that $H_s(t+T)=H_s(t)$, $H_s(t)$ is invariant under $\G$, and $H_{s=0}(t)=H(t)$.
Then, the time-evolution operator of $H_s(t)$, noted as $U_s(t)$, has the expression
\eq{
U_s(t)=\sum_{\bsl{k}} \ket{\psi_{\bsl{k}}} U_s(\bsl{k},t)\bra{\psi_{\bsl{k}}}\ .
}
Further suppose $U_s(\bsl{k},t)$ is a continuous function of $(\bsl{k},t,s)\in\dsR^d\times\dsR\times[0,1]$, and the RGs of $U$ are kept open in the quasienergy band structure given by $U_s(\bsl{k},T)$ as $s$ continuously evolves from $0$ to $1$.
If all the above conditions are satisfied, then (i) after choosing the deformed RGs of $U$ at $s=1$ as the RGs of $U_{s=1}(t)$, then $U_{s=1}$ become a Floquet crystal with $T$, RGs as the deformed RGs of $U$ at $s=1$, and $\G$, and (ii) $U_s(t)$, $\ket{\psi_{\bsl{k},s}}=\ket{\psi_{\bsl{k}}}$ and $T_s=T$ establish the topological equivalence between $U$ and $U_{s=1}$.
The two results follow straightforwardly from the definition of topological equivalence, since the key requirements have been chosen as conditions.

\subsection{OTSL}

In the main text, we presented a formal definition of OTSL.
An equivalent definition for OTSL is the following.
Given a Floquet crystal $\hat{U}$ with $\G$, it has OTSL iff we cannot continuously deform $\hat{U}(t)$ into the time-evolution operator of any static Hamiltonian while keeping all symmetries in $\G$ and keeping open all RGs of $\hat{U}$.
Nevertheless, the formal definition in the main text is more convenient for rigorous derivation, and thereby we will focus on the formal definition. 

The formal definition presented in the main text is based on a formal definition of static limits as discussed in the following.
A static limit is a Floquet crystal with static Hamiltonian.
Given a static Hamiltonian $H_{SL}$, the crystalline symmetry group $\G$ can be straightforwardly determined.
However, $H_{SL}$ is invariant under any time shift, and thereby it does not have a fundamental time period.
Then, in order to define a Floquet crystal based on $H_{SL}$, we need to assign a time period $T_{SL}>0$ for $U_{SL}(t)=\exp(-\ii H_{SL} t )$, and derive the quasienergy band structure and pick the RGs according to $U_{SL}(T_{SL})$.
Then, we have a Floquet crystal $U_{SL}$ characterized by the time-evolution operator $U_{SL}(t)$ equipped with the assigned time period $T_{SL}$, the RG choice according to $U_{SL}(T_{SL})$, and the crystalline symmetry group $\G$.
$U_{SL}$ is a static limit since the underlying Hamiltonian is static.
As the static limit $U_{SL}$ satisfies the definition of Floquet crystals, we can try to establish the topological equivalence between $U_{SL}$ and any other given Floquet crystal.

Clearly, if we choose a different $T_{SL}$ and different RGs for the same time-evolution operator $U_{SL}(t)$, we get a different static limit.
Nevertheless, to determine OTSL for a given Floquet crystal $U$ (with $T$, a RG choice, and $\G$), we do not need to consider $\G$-invariant static limits with $T_{SL}\neq T$.
In other words, if and only if $U$ is topologically distinct from all $\G$-invariant static limits with $T_{SL}= T$, $U$ must have OTSL.
It is because for any $\G$-invariant static limit with $T_{SL}\neq T$, there always exists a topologically equivalent $\G$-invariant static limit with $T_{SL}= T$.
To be more specific, given a static limit $U_{SL}(t)= e^{-\ii H_{SL} t}$ with $T_{SL}$, a relevant gap choice, and $\G$.
We always have another static limit $U_{SL}'(t)= e^{-\ii H_{SL} \frac{T_{SL}}{T} t }$ with $T$, relevant gap choice same as $U_{SL}$, and $\G$.
The same relevant gap choice is allowed by $U_{SL}(T_{SL})=U_{SL}'(T)$, and the topological equivalence is established by $U_s(t)=e^{-\ii H_s t}$ with $s\in [0,1]$, $T_s=(1-s) T_{SL}+s T$, and $H_{s}=H_{SL} T_{SL}/T_s$.
The relevant gaps are preserved by the deformation since $U_s(T_s)=U_{SL}(T_{SL})=U_{SL}'(T)$ for all $s\in [0,1]$ and thereby the quasienergy band structure is independent of $s$.
Other requirements of the topological equivalence can be straightforwardly checked.

\subsection{A Static Limit with Nontrivial Unitary Loop}

At the end of this section, we present a static limit that has a nontrivial unitary loop~\cite{Roy2017FloquetTenFold}.
Let us consider a static Hamiltonian in 0+1D with empty crystalline symmetry group $\G=\emptyset$.
We choose $N=1$, meaning that there is only one basis $\ket{\psi}$ and the static Hamiltonian reads
\eq{
\hat{H}_{SL}=\ket{\psi} h_{SL} \bra{\psi}\ .
}
In this part, we distinguish operator $\hat{O}$ from matrix/scalar $O$.
The representation $h_{SL}$ of Hamiltonian is a scalar, and we set $h_{SL}=2\pi /T$ with $T>0$, resulting in the time-evolution operator as 
\eq{
\hat{U}_{SL}(t)=\ket{\psi} U_{SL}(t) \bra{\psi}
}
with $U_{SL}(t)=e^{-\ii 2\pi t /T}$.
We choose $T$ as the period of the static Hamiltonian.
Then, we have one quasienergy $\E T=0$ within the PBZ $[-\pi,\pi)$, and we only have one quasienergy gap at $-\pi$ (or equivalently $\pi$).
We choose this quasienergy gap as RG, giving us a static limit $U_{SL}$ with assigned period $T$, RG at $\pi$, and $\G=\emptyset$.

According to Sec.\,VA1 of \refcite{Roy2017FloquetTenFold}, in the class-A  case with only one RG, we should first shift the RG to $\pi$ and then deform the time-evolution matrix to a unitary loop to use the K-theoretic classification.
In particular, \refcite{Roy2017FloquetTenFold} shows that the class-A K-theoretic classification of unitary loops for zero spatial dimension is $\dsZ$.
Since the RG is already at $\pi$, we do not need to shift it, and since we already have $U_{SL}(T)=U_{SL}(0)=1$, $U_{SL}(t)$ itself is a unitary loop and we do not need any deformation of it.
As the $\U{1}$ winding of $U_{SL}(t)$ is
\eq{
\frac{1}{2\pi}\int_0^T dt\ U_{SL}^{-1}(t)\ii \partial_t U_{SL}(t) = 1\ ,
}
the unitary loop $U_{SL}(t)$ is nontrivial.
Therefore, the $0+1D$ static limit $U_{SL}$ has a nontrivial unitary loop.

\section{Details on the 2+1D p1 Example}
\label{app:2D_p1}

In this section, we provide more details on the OTSL in the two-band 2+1D dynamical Floquet crystal $U$ with time period $T$ and $\G=p1$,
and on the continuous deformation that establishes the topological equivalence between the 2+1D direct-sum $U_{DS}$ and a 4-band static limit.

\subsection{OTSL in 2+1D Dynamical $U$}

Before going into details, we first discuss the minor differences between \eqTwoPlusOneDDynamicalh in the main text and the two-band dynamical model with anomalous edge modes in \refcite{Rudner2013AFTI}.
To derive \eqTwoPlusOneDDynamicalh from the model in \refcite{Rudner2013AFTI}, we need to (i) perform a unitary transformation $\mat{ 1 & \\ & e^{-\ii k_x}}$ and a momentum transformation $(k_x,k_y)\rightarrow (k_x-k_y,k_x+k_y)/2$, (ii) set $T=2\pi$, $a=1$, $J=1.25$ and $\delta_{AB}=0.1$ according to the convention in \refcite{Rudner2013AFTI}, (iii) omit the hopping in the first segment, and (iv) add an identity term $E_0 \sigma_0$.

Now we discuss the OTSL.
According to \figtwod(a), two bulk RGs separate two bulk quasienergy bands into two isolated sets, where each isolated set only contains one bulk quasienergy band.
We label the isolated sets as $l=1,2$, with the first isolated set lower than the second isolated set.
To determine the OTSL, we use the return map $U_{\Phi}(\bsl{k},t)$ as 
\eq{
U_{\Phi}(\bsl{k},t)=U(\bsl{k},t) [U(\bsl{k},T)]_{\Phi}^{-t/T}
}
with 
\eq{
[U(\bsl{k},T)]_{\Phi}^{-t/T} = \sum_{l=1}^2 \exp\left[  - \frac{t}{T}\log_{\Phi_{\bsl{k}}} e^{-\ii \E_{l,\bsl{k}} T} \right] P_l(\bsl{k},T)\ ,
}
where $\Phi_{\bsl{k}}$ is the PBZ lower bound.
Here $\E_{l,\bsl{k}}$ labels the quasienergy bands derived from $U(\bsl{k},T)$ satisfying $\E_{l,\bsl{k}}T\in[\Phi_{\bsl{k}},\Phi_{\bsl{k}}+2\pi)$ and $\E_{2,\bsl{k}}>\E_{1,\bsl{k}}$, and $P_l(\bsl{k},T)$ is the projection matrix given by the eigenvector of $\E_{l,\bsl{k}}$.
From the return map, we can derive the $\pi_3$ winding number~\cite{Rudner2013AFTI,Yao2017FloquetTenFold} as 
\eq{
\label{eq:W_pi_3}
W_{\pi_3}=\frac{1}{24\pi^2}\int_0^T dt  \int d k_x d k_y \epsilon^{i_1 i_2 i_3}\Tr[U_{\Phi}^\dagger \partial_{i_1} U_{\Phi} U_{\Phi}^\dagger \partial_{i_2} U_{\Phi} U_{\Phi}^\dagger \partial_{i_3} U_{\Phi}]
}
with $(\partial_0,\partial_1,\partial_2)=(\partial_t,\partial_{k_x},\partial_{k_y})$.
Direct calculation shows $W_{\pi_3}=1$ for the 2+1D dynamical $U$ with $\Phi_{\bsl{k}}=-\pi$.
Next, we use $W_{\pi_3}=1$, together with the zero Chern numbers of quasienergy bands, to demonstrate OTSL in $U$.

For any 2+1D $\G$-invariant Floquet crystal $U'$ that is topologically equivalent to $U$, the continuous deformation that establishes the topological equivalence must provide a continuously deformed PBZ (starting from the PBZ of $U$) and thus a continuously deformed return map~\cite{Yu2021DSI}.
Then along the deformation, $W_{\pi_3}$ would stay continuous and thus constant as $W_{\pi_3}\in\dsZ$, meaning that if we choose the deformed PBZ at the end of the deformation as the PBZ of $U'$, $U'$ must have the same $\pi_3$ winding number as $U$.
Moreover, as the continuous deformation always keeps the bulk RGs open, the Chern number of each isolated set must also coincide for $U'$ and $U$ with the PBZ choice that matches the deformation.

Suppose there exists a $\G$-invariant static limit $U_{SL}$ with $T_{SL}=T$ that is topologically equivalent to $U$.
Then, $U_{SL}$ has two bands and two RGs, and there exist a PBZ choice for $U_{SL}$ such that $U_{SL}$ has zero Chern number for each quasienergy band and $W_{\pi_3}^{SL}=1$.

According to \refcite{Yu2021DSI}, as a $\G$-invariant static limit with $T_{SL}=T$, $U_{SL}$ must satisfy
\eqa{
& U_{SL}(t)=e^{-\ii H_{SL} t}\\
& H_{SL}=\sum_{\bsl{k}}\ket{\psi^{SL}_{\bsl{k}}} h_{SL}(\bsl{k})\bra{\psi^{SL}_{\bsl{k}}}\\
& U_{SL}(t)=\sum_{\bsl{k}}\ket{\psi^{SL}_{\bsl{k}}} U_{SL}(\bsl{k},t)\bra{\psi^{SL}_{\bsl{k}}}\\
& U_{SL}(\bsl{k},t)= e^{-\ii h_{SL}(\bsl{k}) t}\ ,
}
where $\ket{\psi^{SL}_{\bsl{k}}}$ labels a choice of bases for $U_{SL}$.
Note that the static Hamiltonian $H_{SL}$ here should not be confused with the static Hamiltonian that is added to the Floquet crystal in order to determine the fragility of OTSL.
Imposing the two-band and two-RG requirement, $h_{SL}(\bsl{k})$ and $U_{SL}(\bsl{k})$ have the form
\eqa{
& h_{SL}(\bsl{k})=\sum_{l=1}^2 E_{l}(\bsl{k}) P_{l}(\bsl{k}) \\
&U_{SL}(\bsl{k},t)=\sum_{l=1}^2 e^{-\ii E_{l}(\bsl{k}) t} P_{l}(\bsl{k})\\
& \E_{l}(\bsl{k}) T=\ii \log_{\Phi_{SL,\bsl{k}}}e^{-\ii E_{l}(\bsl{k}) T} = E_{l}(\bsl{k}) T + j_l 2\pi\ ,
}
where $P_{l}(\bsl{k})$ is the projection matrix for the $E_{l}(\bsl{k})$ band, $\Phi_{SL,\bsl{k}}$ is a generic PBZ lower bound for $U_{SL}$, $\E_{l}(\bsl{k})$ is the quasienergy band, $l$ labels the isolated set in which $\E_{l}(\bsl{k})$ lies, and $j_l\in \dsZ$.
The corresponding return map then reads
\eq{
U_{SL,\Phi_{SL}}(\bsl{k},t) = \sum_{l=1}^2 e^{\ii 2\pi j_l \frac{t}{T}} P_{l}(\bsl{k})\ ,
}
resulting in 
\eq{
\label{eq:W_pi_3_SL}
W_{\pi_3}^{SL}=\sum_{l=1}^2 j_l Ch_l^{SL}
}
with 
\eq{
Ch_l^{SL}=\frac{-\ii}{2\pi}\int dk_x dk_y \Tr[P_l\partial_{k_x} P_l \partial_{k_y} P_l]-(\partial_{k_x}\leftrightarrow \partial_{k_y})
}
the Chern number of the $l$th quasienergy band~\cite{Rudner2013AFTI} of the static limit.
Then, for any $\G$-invariant static limit $U_{SL}$ with $T_{SL}=T$ and two bands and two RGs, choosing the PBZ of $U_{SL}$ to yield zero Chern number for each quasienergy band ($Ch_1^{SL}=Ch_2^{SL}=0$) infers $W_{\pi_3}^{SL}=0$, contradicting to the above-mentioned existence of PBZ for zero Chern numbers and $W_{\pi_3}^{SL}=1$.
Therefore, $U$ cannot be topologically equivalent to any $\G$-invariant static limits with $T_{SL}=T$ and thus has OTSL according to \appref{app:gen_discu}.

We argue that due to $W_{\pi_3}=1\neq 0$, the K-theoretic classification in \refcite{Roy2017FloquetTenFold} identifies the dynamical $U$ as stable dynamical topological.
To see this, note that the K-theoretic classification based on unitary loops in \refcite{Roy2017FloquetTenFold} for symmetry class A can be reproduced by the classification of return maps in \refcite{Yao2017FloquetTenFold}.
So it is legitmate to expect that for the dynamical $U$, its return map is essentially equivalent to its unitary loop in the sense of stable topology.
According to \refcite{Yao2017FloquetTenFold}, $W_{\pi_3}$ is a topological invariant for $2+1$D class-A return maps.
Then, $W_{\pi_3}=1\neq 0$ means the dynamical $U$ has a nontrivial unitary loop.
\refcite{Roy2017FloquetTenFold} defined dynamical topological systems by nontrivial unitary loops, and thereby the dynamical $U$ is dynamical topological according to \refcite{Roy2017FloquetTenFold}.
As K-theory can only identify stable topology, the dynamical $U$ is stable dynamical topological according to \refcite{Roy2017FloquetTenFold}.

Having nontrivial unitary loops is not equivalent to having OTSL.
As exemplified in \appref{app:gen_discu}, static limits may still have nontrivial unitary loops.
In terms of $\pi_3$ winding number, both \eqnref{eq:W_pi_3_SL} and Appendix C in \refcite{Rudner2013AFTI} suggest that 2+1D static limits with nonzero Chern numbers may have nonzero $\pi_3$ winding numbers.
Therefore, being stable dynamical topological according to \refcite{Roy2017FloquetTenFold} does not prevent the dynamical $U$ from having fragile OTSL.

\subsection{Continuous Deformation Connecting $U_{DS}$ and a 4-Band Static Limit}
\label{app:2+1D_P_Deform}

As discussed in the main text, we deform $U_{DS}$ into to a four band static limit through $\widetilde{U}_{2D,s_1}(t)$.

To construct $\widetilde{U}_{2D,s_1}(t)$, we first consider 
\eq{
\label{eq:2+1D_P_Ht}
\widetilde{H}_{2D}(t;s_1,s_2)=\sum_{\bsl{k}}\ket{\psi^{DS}_{\bsl{k}}} \widetilde{h}_{2D}(\bsl{k},t;\delta,s_1,s_2) \bra{\psi^{DS}_{\bsl{k}}}\ ,
}
where
\eq{
\widetilde{h}_{2D}(\bsl{k},t;\delta,s_1,s_2)=
\mat{
0.01+ \sigma_x D_x(\bsl{k},t,s_1) + \sigma_y D_y(\bsl{k},t,s_1)+\sigma_z (\delta-\frac{s_1}{2\pi}\sin(k_y)) & 0.2 s_2 -\ii 0.2 s_2 \sigma_y\\
0.2 s_2 + \ii 0.2 s_2 \sigma_y & 
\frac{1}{2}+\frac{(M(\bsl{k})-2s_1)\sigma_z+\sin(k_x)\sigma_x+\sin(k_y)\sigma_y}{(3+s_1)\pi}
}
}
\eqa{
& D_x(\bsl{k},t,s_1)=-\frac{s_1}{2\pi}(-0.5+\cos(k_x)+\cos(k_y))+\widetilde{d}_x(\bsl{k},t,s_1)\\
& D_y(\bsl{k},t,s_1)=-\frac{s_1}{2\pi}\sin(k_x)+\widetilde{d}_y(\bsl{k},t,s_1)\ ,
}
and $s_1,s_2\in[0,1]$.
The expressions of $\widetilde{d}_{x,y}$ are shown in \tabref{tab:2+1D_P_para_deform_app}, and $\widetilde{H}_{2D}(t;\delta,s_1,s_2)=\widetilde{H}_{2D}(t+T;\delta,s_1,s_2)$ is imposed.
By defining $\widetilde{U}_{2D}(t;\delta,s_1,s_2)$ as the time evolution operator of $\widetilde{H}_{2D}(t;\delta,s_1,s_2)$, we have $\widetilde{U}_{2D}(t;0.1,0,0)=U_{DS}(t)$, and $\widetilde{H}_{2D}(t;0,1,0)$ is static.
The deformation is defined as $\widetilde{U}_{2D,s_1}(t)=\widetilde{U}_{2D}(t;0.1(1-s_1),s_1,0.7\sin(s_1\pi))$.
As shown in \figtwod(d), the deformation does not close any of the RGs for $s_1\in[0,1]$.
Then according to the argument in \appref{app:gen_discu}, we can choose the deformed RGs at $s_1=1$ as the RGs of $\widetilde{U}_{2D,1}$, and $\widetilde{U}_{2D,1}$ becomes the four-band static limit discussed in the main text since $\widetilde{H}_{2D}(t;0,1,0)$ is static.
Moreover, $\widetilde{U}_{2D,1}$ is topologically equivalent to $U_{DS}$, meaning that $U_{DS}$ has no OTSL.

\begin{table}[t]
    \centering
    \begin{tabular}{c|c|c|c|c|c}
        $t$ & $[0, T/5)$ & $[T/5, 2T/5)$ & $[2T/5, 3T/5)$ & $[3T/5, 4T/5)$ & $[4T/5, T)$\\
         \hline
         $\widetilde{d}_{x}$ &  0 & $-1.25(1-s_1)\cos(k_y)$ & $-1.25(1-s_1)\cos(k_x-k_y)$ & $-1.25(1-s_1)\cos(k_x)$ & 0 \\ $\widetilde{d}_{y}$ &  0 & $1.25(1-s_1)\sin(k_y)$ & $-1.25(1-s_1)\sin(k_x-k_y)$ & $-1.25(1-s_1)\sin(k_x)$ & 0 \\
    \end{tabular}
    \caption{The expressions of $\widetilde{d}_{x,y}$ \eqnref{eq:2+1D_P_Ht} for $t\in [0,T)$.}
    \label{tab:2+1D_P_para_deform_app}
\end{table}

\section{A 1+1D Inversion-Invariant Model}
\label{app:1+1D_P}

\begin{figure}[h]
    \centering
    \includegraphics[width=0.5\columnwidth]{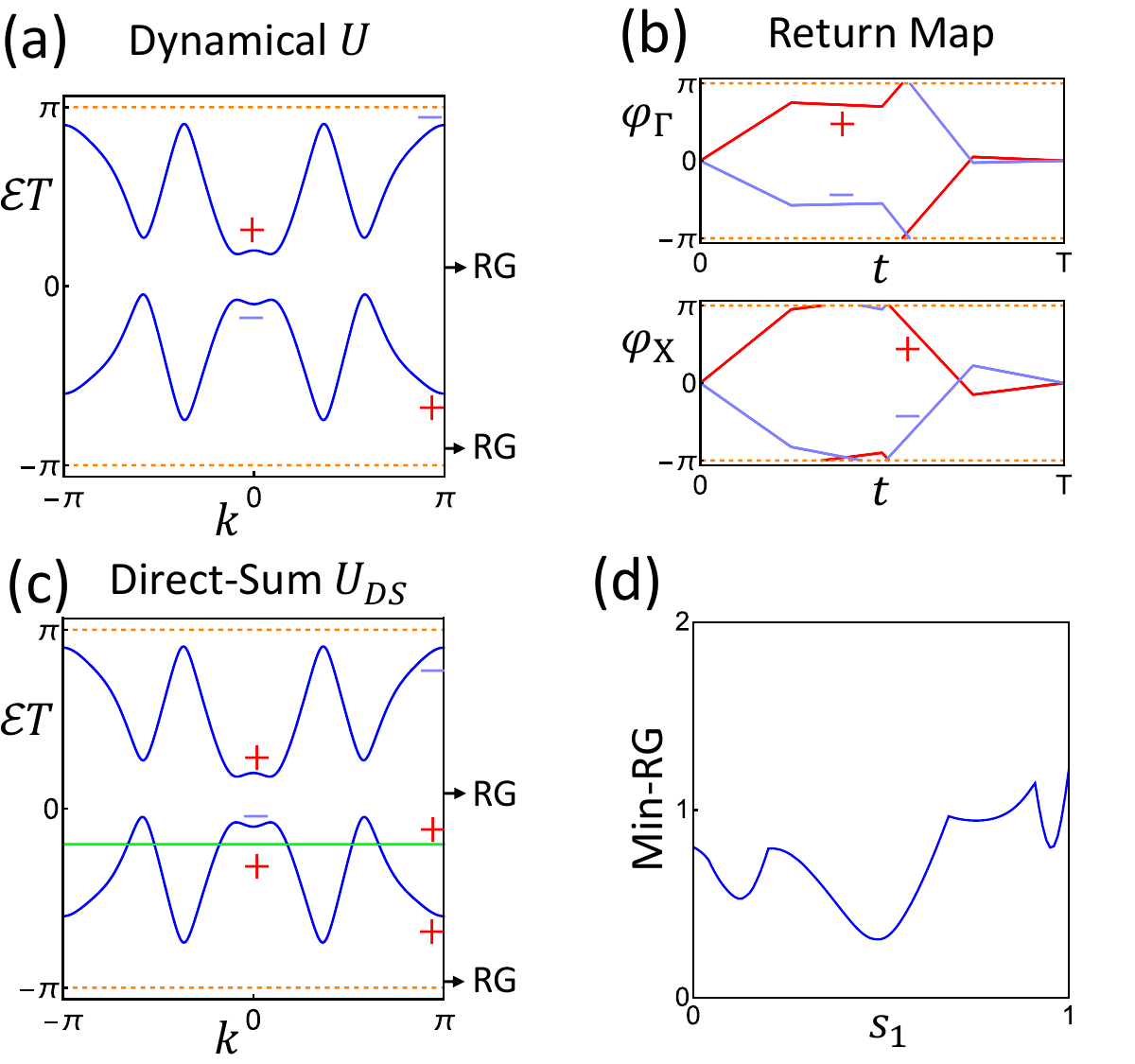}
    \caption{In this figure, we show the dynamical fragile topology in the 1+1D inversion-invariant example.
    In (a), we show the quasienergy bands (blue) for the 1+1D Floquet crystal $U$ that has OTSL. 
    In (b), we plot the phase bands of the return map for $U$ at $\Gamma$ and $X$.
    In (c), we plot the quasienergy bands of the direct-sum Floquet crystal $U_{DS}$.
    The green line stands for the quasienergy band brought by the added static Hamiltonian $H_{SL}$, while the blue lines are given by $U$.
    In (a-c), $\pm$ stand for the parities of the eigenvectors, and the orange dashed lines are the PBZ boundaries.
    In (d), we plot the minimum of the deformed RGs for the deformation $\widetilde{U}_{1D,s_1}$ that connects $U_{DS}$ ($s_1=0$) to a three-band static limit ($s_1=1$).
    }
    \label{fig:1+1D_P}
\end{figure}

In this part, we introduce a 1+1D example with $\G$ spanned by 1D lattice translations and inversion.
We consider a simple 1D lattice and put two spinless orbitals (one $\s$ orbital and one $\p$ orbital) at each site, resulting in the bases $\ket{\psi_k} = (\ket{\psi_{k,\s}}, \ket{\psi_{k,\p}})$.
In this (and next) example, we always set the lattice constant to be $1$.
We construct the dynamical $T$-periodic Hamiltonian $H(t)$ by including time-dependent onsite energy $E_{\s}, E_{\p}$ and nearest-neighboring hopping $J_{\s}, J_{\p}, J_{\s\p}$; then, the matrix representation of $H(t)$ reads
\eq{
\label{eq:1+1D_P_h}
h(k,t)= 
\mat{
E_{\s}(t) + J_{\s}(t) \cos(k) & J_{\s\p}(t) \sin(k) \\
J_{\s\p}^*(t) \sin(k) & E_{\p}(t) + J_{\p}(t) \cos(k)
}\ .
}
(See detailed expressions in \tabref{tab:1+1D_P_para_formaintext_app}.)
The 1+1D $H(t)$ preserves the inversion symmetry $\mathcal{P}$ as $\mathcal{P}\ket{\psi_k} = \ket{\psi_{-k}} \sigma_z$ and $\sigma_z h(k,t) \sigma_z^\dagger = h(-k,t)$, where $\sigma_z$ is the $z$ Pauli matrix.
Combining \eqnref{eq:1+1D_P_h} with \eqUfromHgen, we can derive the time-evolution operator $U(t)$ and the time-evolution matrix $U(k,t)$ in the space spanned by $\ket{\psi_k}$.
In \figref{fig:1+1D_P}(a), we plot the quasienergy bands in $[-\pi,\pi)$ by diagonalizing $U(k,T)$, showing two quasienergy gaps, both of which we choose to be RGs.
Then, we have a Floquet crystal $U$ characterized by the 1+1D $U(t)$ equipped with $T$, the RG choice \figref{fig:1+1D_P}(a), and $\G$.
Since $\Phi_k=-\pi$ lies in a RG, $[-\pi,\pi)$ is a PBZ of $U$.

To determine the OTSL of $U$, we derive the time-periodic return map~\cite{Nathan2015TopoSingClas,Yao2017FloquetTenFold} $U_{\Phi}(k,t)$ from $U(k,t)$, and plot the phase bands of $U_{\Phi}(k,t)$ at two inversion-invariant momenta $\Gamma(k=0)$ and $X(k=\pi)$ in \figref{fig:1+1D_P}(b).
Since $U_{\Phi}(k,t)$ preserves the inversion symmetry, eigenvectors of each phase band at $k_0=\Gamma/X$ have a definite parity $\alpha=\pm$.
We can then count the total winding of the phase bands with parity $\alpha$ at $k_0$, noted as $\nu_{k_0,\alpha}$, yielding $(\nu_{\Gamma,+},\nu_{\Gamma,-},\nu_{X,+},\nu_{X,-})=(1,-1,0,0)$ for \figref{fig:1+1D_P}(b).
In particular, $U$ has a nonzero difference between $\nu_{\Gamma,+}$ and $\nu_{X,-}$, namely $\nu_{\Gamma,+}-\nu_{X,-}=1$.
On the contrast, any $\G$-invariant static limit with the parity distribution like \figref{fig:1+1D_P}(a) must have $\nu_{\Gamma,+}-\nu_{X,-}=0$.
It is because the winding numbers of those static limits are given by winding each band independently along time; since $\nu_{\Gamma,+}$ and $\nu_{X,-}$ are given by two eigenvectors in the same band, they must be the same for those static limits.
As a result, $\nu_{\Gamma,+}-\nu_{X,-}=1$ cannot be reproduced by any of those static limits, indicating that $U$ has OTSL.
In fact, $(\nu_{\Gamma,+}-\nu_{X,-})$ is nothing but the dynamical symmetry indicator (DSI) proposed in \refcite{Yu2021DSI}. (See \appref{sec:1+1D_P_DSI} for details.)

To show the fragility of OTSL in the dynamical $U$, we need to add a static Hamiltonian to form a direct-sum Floquet crystal.
For this purpose, we consider a static 1D chain of $\dd$ orbitals, whose static Hamiltonian reads $H_{SL}=\sum_{k} \ket{\psi_{k,\dd}} E_{\dd} \bra{\psi_{k,\dd}}$ with $\ket{\psi_{k,\dd}}$ the bases.
$H_{SL}$ preserves $\G$ owing to $\mathcal{P}\ket{\psi_{k,\dd}}=\ket{\psi_{-k,\dd}}$, and the addition $H_{DS}(t)=H(t)+H_{SL}$ is in the direct-sum way as $\ket{\psi_{k,\dd}}$ are orthogonal to the bases for $U$.
By choosing $\ket{\psi^{DS}_k}=(\ket{\psi_{k,\s}},\ket{\psi_{k,\p}},\ket{\psi_{k,\dd}})$, the direct-sum time-evolution operator $U_{DS}(t)$ is represented by $U_{DS}(k,t)$ in \eqDSUmatgen with $h_{SL}(k)=E_{\dd}$.
We plot the quasienergy bands given by $U_{DS}(k,T)$ in \figref{fig:1+1D_P}(c), showing that both the RGs of the dynamical $U$ are preserved.
Therefore, we have a direct-sum Floquet crystal $U_{DS}$ with $T$, the RG choice same as $U$ (\figref{fig:1+1D_P}(c)), and $\G$.

To sufficiently demonstrate that $U_{DS}$ has no OTSL, we explicitly construct a symmetry-preserving continuous deformation $\widetilde{U}_{1D,s_1}(t)$ with a tuning parameter $s_1\in[0,1]$, such that $\widetilde{U}_{1D,0}(t)=U_{DS}(t)$ and $\widetilde{U}_{1D,1}(t)$ corresponds to a three-band static limit. 
As shown by \figref{fig:1+1D_P}(d), the deformation does not close any RG, establishing the topological equivalence between the direct-sum $U_{DS}$ and a static limit.
(See more details in \appref{app:1+1D_P_Deform}).
Therefore, the direct-sum $U_{DS}$ has no OTSL, and the OTSL in the 1+1D dynamical $U$ is fragile.

In the following, we will provide more details on the calculation of DSI of the 1+1D dynamical $U$ and on the deformation $\widetilde{U}_{1D,s_1}(t)$.

\begin{table}[t]
    \centering
    \begin{tabular}{c|cccc}
        $t$ & $[0, T/4)$ & $[T/4, 2T/4)$ & $[2T/4, 3T/4)$ & $[3T/4, T)$\\
         \hline
         $E_{\s}$ &  $1.6$ & $0$ & $0$ & $0$ \\ 
         $E_{\p}$ &  $-1.2$ & $0$ & $0$ & $0$ \\ 
         $J_{\s}$ &  $0$ & $0$ & $2.8$ & $0$ \\
         $J_{\p}$ &  $0$ & $0$ & $-3$ & $0$ \\ 
         $J_{\s\p}$ & $0$ & $-0.4 -\ii 0.6$ & $0$ & $-0.4 -\ii 0.6$\\
         $E_{\dd}$ & $-1.1$ & $-1.1$ & $-1.1$ & $-1.1$\\
    \end{tabular}
    \caption{The expressions of $E_{\s}$, $J_{\s}$, $J_{\s\p}$, $E_{\p}$, $J_{\p}$, and $E_{d}$ for the 1+1D $U$ and $U_{DS}$ within one time period.
    All parameters are invariant under $t\rightarrow t+T$.
    Besides, we choose $T=2\pi$ throughout this section.
    }
    \label{tab:1+1D_P_para_formaintext_app}
\end{table}

\subsection{DSI of 1+1D dynamical $U$}

\label{sec:1+1D_P_DSI}

In this part, we calculate the DSI of $U$ following \refcite{Yu2021DSI}.
To do this, we need to first derive the symmetry and winding data for $U$.
According to \figref{fig:1+1D_P}(a), two RGs separate two quasienergy bands into two isolated sets, where each isolated set only contains one quasienergy band.
We label the isolated sets as $l=1,2$, with the first isolated set lower than the second isolated set in \figref{fig:1+1D_P}(a).
As $\G$ is spanned by the inversion symmetry and 1D lattice translations, the symmetry data is determined by the number of eigenvectors with each parity ($\alpha=\pm$) at each inversion-invariant momentum ($k_0=\Gamma$ or $X$) in each isolated set of the quasienergy bands, noted as $n^{l}_{k_0,\alpha}$.
Then, we have a four-component vector $A_l = (n_{\Gamma,+}^l,n_{\Gamma,-}^l,n_{X,+}^l,n_{X,-}^l)^T$ for the $l$th IS, and the symmetry data can be read out from \figref{fig:1+1D_P}(a) as 
\eq{
\label{eq:1+1D_P_SD}
A=(A_1\ A_2)\text{ with }A_1= (0,1,1,0)^T\text{ and }A_2=(1,0,0,1)^T\ .
}

The winding data is derived from the time-periodic return map $U_{\Phi}(k,t)$, which is defined as 
\eq{
\label{eq:1+1D_P_RM}
U_{\Phi}(k,t)=U(k,t) [U(k,T)]_{\Phi}^{-t/T}
}
with 
\eq{
[U(k,T)]_{\Phi}^{-t/T} = \sum_{l=1}^2 \exp\left[  - \frac{t}{T}\log_{\Phi_k} e^{-\ii \E_{l,k} T} \right] P_l(k,T)\ .
}
Here $\E_{l,k}$ labels the quasienergy bands derived from $U(k,T)$ satisfying $\E_{l,k}T\in[\Phi_k,\Phi_k+2\pi)$ with $\Phi_k=-\pi$, and $P_l(k,T)$ is the projection matrix given by the eigenvector of $\E_{l,k}$.
The logarithm in the above equation is defined as $\ii \log_{\Phi_k}e^{-\ii x}\in [\Phi_k,\Phi_k+2\pi)$ for all $x\in\dsR$, resulting in
\eq{
\ii \log_{\Phi_k}e^{-\ii \E_{l,k} T}= \E_{l,k} T
}
since we have chosen $\E_{l,k} T\in [\Phi_k,\Phi_k+2\pi)$.
The return map preserves the inversion symmetry as $\sigma_z U_{\Phi}(k,t) \sigma_z =U_{\Phi}(-k,t)$, yielding
\eq{
U_{\Phi}(k_0,t)=\mat{U_{\Phi,+}(k_0,t) & \\ &  U_{\Phi,-}(k_0,t)}\ ,
}
where $U_{\Phi,\pm}(k_0,t)$ are the parity $\pm$ blocks of the block-diagonalized $U_{\Phi}(k_0,t)$.
Then, the winding data reads $V=(\nu_{\Gamma,+},\nu_{\Gamma,-},\nu_{X,+},\nu_{X,-})^T$, where
\eq{
\label{eq:1+1D_P_WD_gen}
\nu_{k_0,\alpha}= \frac{1}{2\pi}\int_0^T dt \Tr[U_{\Phi,\alpha}^\dagger(k_0,t)\ii\partial_t U_{\Phi,\alpha}(k_0,t)]\ .
}

To calculate $V$, first note 
\eq{
h(k_0,t)=\mat{E_{\s}(t)+J_{\s}(t) \cos(k_0)  & \\ & E_{\p}(t)+J_{\p}(t) \cos(k_0) }\ ,
}
and then we have
\eq{
U(k_0,t)=\mat{ 
\exp\left[-\ii  f_{\s}(k_0,t) \right]& \\
 & \exp\left[-\ii f_{\p}(k_0,t) \right]
}\ ,
}
where $f_{a}(k_0,t)=\int_0^t dt' [E_{a}(t')+J_{a}(t') \cos(k_0)]$ with $a=\s,\p$.
Then, the return map reads
\eq{
U_{\Phi}(k_0,t)=\mat{ 
\exp\left[-\ii  f_{\s}(k_0,t) -\frac{t}{T}\log_{\Phi_{k_0}} e^{-\ii f_{\s}(k_0,T)}\right]& \\
 & \exp\left[-\ii f_{\p}(k_0,t) -\frac{t}{T}\log_{\Phi_{k_0}} e^{-\ii f_{\p}(k_0,T)}\right]
}\ ,
}
and combined with \eqnref{eq:1+1D_P_WD_gen}, we have 
\eqa{
& \nu_{k_0,+}=\frac{1}{2\pi}[f_{\s}(k_0,T) - \ii\log_{\Phi_{k_0}} e^{-\ii f_{\s}(k_0,T)}] \\
& \nu_{k_0,-}=\frac{1}{2\pi}[f_{\p}(k_0,T) - \ii\log_{\Phi_{k_0}} e^{-\ii f_{\p}(k_0,T)}]\ .
}
Using \tabref{tab:1+1D_P_para_formaintext_app}, we arrive at
\eq{
\label{eq:1+1D_P_WD}
V=(1,-1,0,0)^T\ .
}

In general, the winding data for 1+1D inversion-invariant Floquet crystals must satisfy a compatibility relation~\cite{Yu2021DSI} $\nu_{\Gamma,+}+\nu_{\Gamma,-}=\nu_{X,+}+\nu_{X,-}$, since the total winding at different momenta should be the same.
Let us now consider all Floquet crystals with symmetry data equivalent to $U$, as inequivalent symmetry data must infer topological distinction.
Here having symmetry data equivalent to $U$ means that having the same symmetry data as \eqnref{eq:1+1D_P_SD} upon choosing PBZs.
For this specific example, having symmetry data equivalent to $U$ simply means that having symmetry data with $A_1$ and $A_2$ in \eqnref{eq:1+1D_P_SD} as its two columns.
When certain symmetry representations are missing, the corresponding winding numbers must be zero, giving an extra constraint on the winding data.
Since those Floquet crystals have eigenvectors with both parities at both inversion-invariant momenta, there is no extra constraint due to the missing representations, and then their winding data take value in 
\eq{
\{ V\}= \{(\nu_{\Gamma,+},\nu_{\Gamma,-},\nu_{\Gamma,+}+\nu_{\Gamma,-}-\nu_{X,-},\nu_{X,-})^T|\nu_{\Gamma,+},\nu_{\Gamma,-},\nu_{X,-}\in\dsZ \}\ .
}

On the other hand, the winding data of $\G$-invariant static limits with $T_{SL}=T$ and symmetry data equivalent to $U$ always belong to 
\eq{
\label{eq:V_SL_set}
\{ V_{SL}\}=\{ \sum_{l} q_l A_l=(q_2,q_1,q_1,q_2)^T | q_1,q_2\in\dsZ\}\ ,
}
which is given by winding each connected isolated set independently.
To be more specific, let us consider a generic $\G$-invariant static limit $U_{SL}$ with $T_{SL}=T$ and symmetry data equivalent to $U$.
$U_{SL}$ must satisfy
\eqa{
& U_{SL}(t)=e^{-\ii H_{SL} t}\\
& H_{SL}=\sum_{k}\ket{\psi^{SL}_{k}} h_{SL}(k)\bra{\psi^{SL}_{k}}\\
& U_{SL}(t)=\sum_{k}\ket{\psi^{SL}_{k}} U_{SL}(k,t)\bra{\psi^{SL}_{k}}\\
& U_{SL}(k,t)= e^{-\ii h_{SL}(k) t}\ ,
}
where $\ket{\psi^{SL}_{k}}$ labels a choice of bases for $U_{SL}$.
Note that the static Hamiltonian $H_{SL}$ here should not be confused with the static Hamiltonian that is added to the Floquet crystal in order to determine the fragility of OTSL.
Owing to the equivalent symmetry data, $U_{SL}$ must have the same number of bands and the same number of RGs as $U$.
Then, we impose the two-band and two-RG requirement, and get the following form of $h_{SL}(k)$ and $U_{SL}(k)$
\eqa{
& h_{SL}(k)=\sum_{l=1}^2 E_{l}(k) P_{l}(k) \\
&U_{SL}(k,t)=\sum_{l=1}^2 e^{-\ii E_{l}(k) t} P_{l}(k)\\
& \E_{l}(k) T=\ii \log_{\Phi_{SL,k}}e^{-\ii E_{l}(k) T} = E_{l}(k) T - q_l 2\pi\ ,
}
where $P_{l}(k)=Y_l(k) Y^\dagger_l(k)$, $Y_l(k)$ is the orthonormal eigenvector of $h_{SL}(k)$ for the $E_{l}(k)$ band, $\Phi_{SL,k}$ is a generic PBZ lower bound for $U_{SL}$, $\E_{l}(k)$ is the quasienergy band, $l$ labels the isolated set in which $\E_{l}(k)$ lies, and $q_l\in \dsZ$.
The corresponding return map then reads
\eq{
U_{SL,\Phi_{SL}}(k,t) = \sum_{l=1}^l e^{-\ii 2\pi q_l \frac{t}{T}} P_{l}(k)\ .
}
As we require the static limit to have symmetry data equivalent to $U$, $P_{l}(k)$'s should reproduce $A_1$ and $A_2$ in \eqnref{eq:1+1D_P_SD}.
Without loss of generality, we can choose $P_{l}(k)$ to reproduce $A_l$, respectively, meaning that $Y_{1}(0)$ has parity $-$, $Y_{1}(\pi)$ has parity $+$, $Y_{2}(0)$ has parity $+$, and $Y_{2}(\pi)$ has parity $-$.
With the eigenvectors as the bases, we can block diagonalize $U_{SL,\Phi_{SL}}(k_0,t)$ with $k_0=0,\pi$ to into blocks with definite parities.
Explicitly, we choose $W_{0}= (Y_{2}(0)\  Y_{1}(0))$ and  $W_{\pi}= (Y_{1}(\pi)\  Y_{2}(\pi))$, and then
\eq{
W_{k_0}^\dagger U_{SL,\Phi_{SL}}(k_0,t) W_{k_0} = \mat{ U_{SL,\Phi_{SL},+}(k_0,t) & \\ &  U_{SL,\Phi_{SL},-}(k_0,t) }
}
with
\eqa{
& U_{SL,\Phi_{SL},+}(0,t) = e^{-\ii 2\pi q_2 \frac{t}{T}}\\
& U_{SL,\Phi_{SL},-}(0,t) = e^{-\ii 2\pi q_1 \frac{t}{T}}\\
& U_{SL,\Phi_{SL},+}(\pi,t) = e^{-\ii 2\pi q_1 \frac{t}{T}}\\
& U_{SL,\Phi_{SL},-}(\pi,t) = e^{-\ii 2\pi q_2 \frac{t}{T}}\ .
}
Combined with \eqnref{eq:1+1D_P_WD_gen}, we arrive at $(q_2, q_1, q_1 ,q_2)^T $ as the form of the winding data of $\G$-invariant static limits with $T_{SL}=T$ and symmetry data equivalent to $U$, resulting in \eqnref{eq:V_SL_set}.

If the winding data $V$ in \eqnref{eq:1+1D_P_WD} of the Floquet crystal $U$ does not belong to $\{ V_{SL}\}$, it means that static limits with $T_{SL}=T$ and symmetry data equivalent to $U$ cannot reproduce the winding data of $U$, indicating that $U$ has obstruction to static limits.
The we can define the DSI to take value from the following quotient group
\eq{
X=\frac{\{ V\}}{\{ V_{SL}\}}\approx \{ \nu_{\Gamma,+}-\nu_{X,-} \in\dsZ\}
}
with the expression of DSI being $\nu_{\Gamma,+}-\nu_{X,-}$.
Nonzero DSI infers that $V\notin \{ V_{SL}\}$, and thereby infers OTSL.
Combined with \eqnref{eq:1+1D_P_WD}, we have $\nu_{\Gamma,+}-\nu_{X,-}=1$, which is nonzero and thereby sufficiently infers OTSL.

\subsection{Continuous Deformation Connecting $U_{DS}(t)$ to Static Limit}
\label{app:1+1D_P_Deform}

\begin{table}[t]
    \centering
    \begin{tabular}{c|cccc}
        $t$ & $[0, T/4)$ & $[T/4, 2T/4)$ & $[2T/4, 3T/4)$ & $[3T/4, T)$\\
         \hline
         $\widetilde{E}_{\s}$ &  $1.6 - 2 s_1 $ & $-0.4 s_1$ & $-0.4 s_1$ & $-0.4 s_2$ \\ 
         $\widetilde{E}_{\p}$ &  $-\frac{1.2}{3 s_1+1}$ & $-\frac{1.2}{3 s_1+1} s_1$ & $-\frac{1.2}{3 s_1+1} s_1$ & $-\frac{1.2}{3 s_1+1} s_1$ \\ 
         $\widetilde{J}_{\s}$ &  $0.7 s_1$ & $0.7 s_1$ & $2.8-2.1 s_1$ & $0.7 s_2$ \\
         $\widetilde{J}_{\p}$ &  $-\frac{3}{3 s_1+1}s_1$ & $-\frac{3}{3 s_1+1} s_1$ & $-\frac{3}{3 s_1+1}$ & $-\frac{3}{3 s_1+1} s_1$ \\ 
         $\widetilde{J}_{\s\p}$ & $(-0.4 -\ii 0.6)s_1$ & $-0.4 -\ii 0.6$ & $(-0.4 -\ii 0.6)s_1$ & $-0.4 -\ii 0.6$\\
         $\widetilde{E}_{\dd}$ & $-1.1+0.7 s_1$ & $-1.1+0.7 s_1$ & $-1.1+0.7 s_1$ & $-1.1+0.7 s_1$\\
         $\Lambda_{\s\dd}$ & $0.2 s_1$ & $0.2 s_1$ & $0.2 s_1$ & $0.2 s_1+5s_1 (1-s_2)$ \\
         $J_{\s\dd}$ & $0$ & $0$ & $0$ & $-1.6 s_1 (1-s_2)$ \\
    \end{tabular}
    \caption{The expressions of $\widetilde{E}_{\s}$, $\widetilde{J}_{\s}$, $\widetilde{J}_{\s\p}$, $\Lambda_{\s\dd}$, $J_{\s\dd}$, $\widetilde{E}_{\p}$, $\widetilde{J}_{\p}$, and $\widetilde{E}_{d}$ in \eqnref{eq:1+1D_P_Htilde} for $t\in [0,T)$.}
    \label{tab:1+1D_P_para_deform_app}
\end{table}

In this part, we elaborate on the continuous deformation $\widetilde{U}_{1D,s_1}(t)$ that establishes the topological equivalence between $U_{DS}$ and a static limit.

We first construct a deformation $\widetilde{U}_{1D}(t;s_1,s_2)$ with two tuning parameters $s_1,s_2\in[0,1]$ from the following Hamiltonian 
\eq{
\label{eq:1+1D_P_Htilde}
\widetilde{H}_{1D}(t;s_1,s_2)=\sum_{k} \ket{\psi_{k}^{DS}} 
\widetilde{h}_{1D}(k,t;s_1,s_2)
\bra{\psi_{k}^{DS}}\ ,
}
where
\eqa{
& \widetilde{h}_{1D}(k,t;s_1,s_2)=\\
&\mat{
\widetilde{E}_{\s}(t;s_1,s_2) + \widetilde{J}_{\s}(t;s_1,s_2) \cos(k) & \widetilde{J}_{\s\p}(t;s_1,s_2) \sin(k) & \Lambda_{\s\dd}(t;s_1,s_2)+ J_{\s\dd}(t;s_1,s_2)\cos(k)\\
\widetilde{J}_{\s\p}^*(t;s_1,s_2)  \sin(k) & \widetilde{E}_{\p}(t;s_1,s_2) + \widetilde{J}_{\p}(t;s_1,s_2) \cos(k) & 0 \\
\Lambda_{\s\dd}(t;s_1,s_2)+ J_{\s\dd}(t;s_1,s_2)\cos(k) & 0 & \widetilde{E}_{\dd}(s_1)
}\ ,
}
$\widetilde{H}_{1D}(t+T;s_1,s_2)=\widetilde{H}_{1D}(t;s_1,s_2)$ is imposed, and the expressions of $\widetilde{E}_{\s}$, $\widetilde{J}_{\s}$, $\widetilde{J}_{\s\p}$, $\Lambda_{\s\dd}$, $J_{\s\dd}$, $\widetilde{E}_{\p}$, $\widetilde{J}_{\p}$, and $\widetilde{E}_{\dd}$ are shown in \tabref{tab:1+1D_P_para_deform_app}.
$\widetilde{U}_{1D}(t;s_1,s_2)$ is the time-evolution operator of $\widetilde{H}_{1D}(t;s_1,s_2)$ as
\eq{
\label{eq:1+1D_P_Utilde}
\widetilde{U}_{1D}(t;s_1,s_2) = \sum_{k} \ket{\psi_{k}^{DS}} 
\widetilde{U}_{1D}(k,t;s_1,s_2)
\bra{\psi_{k}^{DS}}\ ,
}
where
\eq{
\widetilde{U}_{1D}(k,t;s_1,s_2)=\mathcal{T}\exp\left[-\ii \int_0^t dt' \widetilde{h}_{1D}(k,t';s_1,s_2) \right]
}
is a continuous function of $(k,t,s_1,s_2)\in \dsR\times\dsR\times[0,1]\times[0,1]$.
Based on
\eq{
\mathcal{P}\ket{\psi_{k}^{DS}}=\ket{\psi_{-k}^{DS}}\mat{ 1 & & \\ & -1 & \\ & & 1}\ ,
}
we know $[\mathcal{P},\widetilde{H}_{1D}(t;s_1,s_2)]=0$ and thus $[g, \widetilde{U}_{1D}(t;s_1,s_2) ]=0$ for all $g\in\G$.
According to \tabref{tab:1+1D_P_para_deform_app}, $\widetilde{H}_{1D}(t;0,0)=H_{DS}(t)$, $\widetilde{U}_{1D}(t;0,0)=U_{DS}(t)$, and $\widetilde{H}_{1D}(t;1,1)$ is static.

We now define Hermitian $H_{1D,s_1}(t)=\widetilde{H}_{1D}(t;s_1,s_1-0.2\sin(\pi s_1))$ and the deformation used in the main text is just $\widetilde{U}_{1D,s_1}(t)=\widetilde{U}_{1D}(t;s_1,s_1-0.2\sin(\pi s_1))$ with $s_1\in[0,1]$.
As $\widetilde{U}_{1D,0}(t)=U_{DS}(t)$ whose RGs have been chosen, we can calculate the quasienergy bands given by $\widetilde{U}_{1D,s_1}(T)$, track the deformed RGs of $U_{DS}$ as varying $s_1$ continuously, and plot the minimum of them in \figref{fig:1+1D_P}(d).
To be more specific, we explicitly plot in \figref{fig:1D_Inversion_app} quasienergy bands given by $\widetilde{U}_{1D,s_1}(T)$, where we can see both the deformed RGs of $U_{DS}$ are kept open.
Then according to \appref{app:gen_discu}, we can choose the deformed RGs of $U_{DS}$ at $s_1=1$ as the RGs of $\widetilde{U}_{1D,1}$, resulting in a Floquet crystal $\widetilde{U}_{1D,1}$ topologically equivalent to $U_{DS}$.
Combined with the fact that $\widetilde{H}_{1D,1}(t)=\widetilde{H}_{1D}(t;1,1)$ is static, we know  $\widetilde{U}_{1D,1}$ is a static limit, and $U_{DS}$ is topologically equivalent to it.
Then, by definition, $U_{DS}$ has no OTSL.

\begin{figure}[t]
    \centering
    \includegraphics[width=\columnwidth]{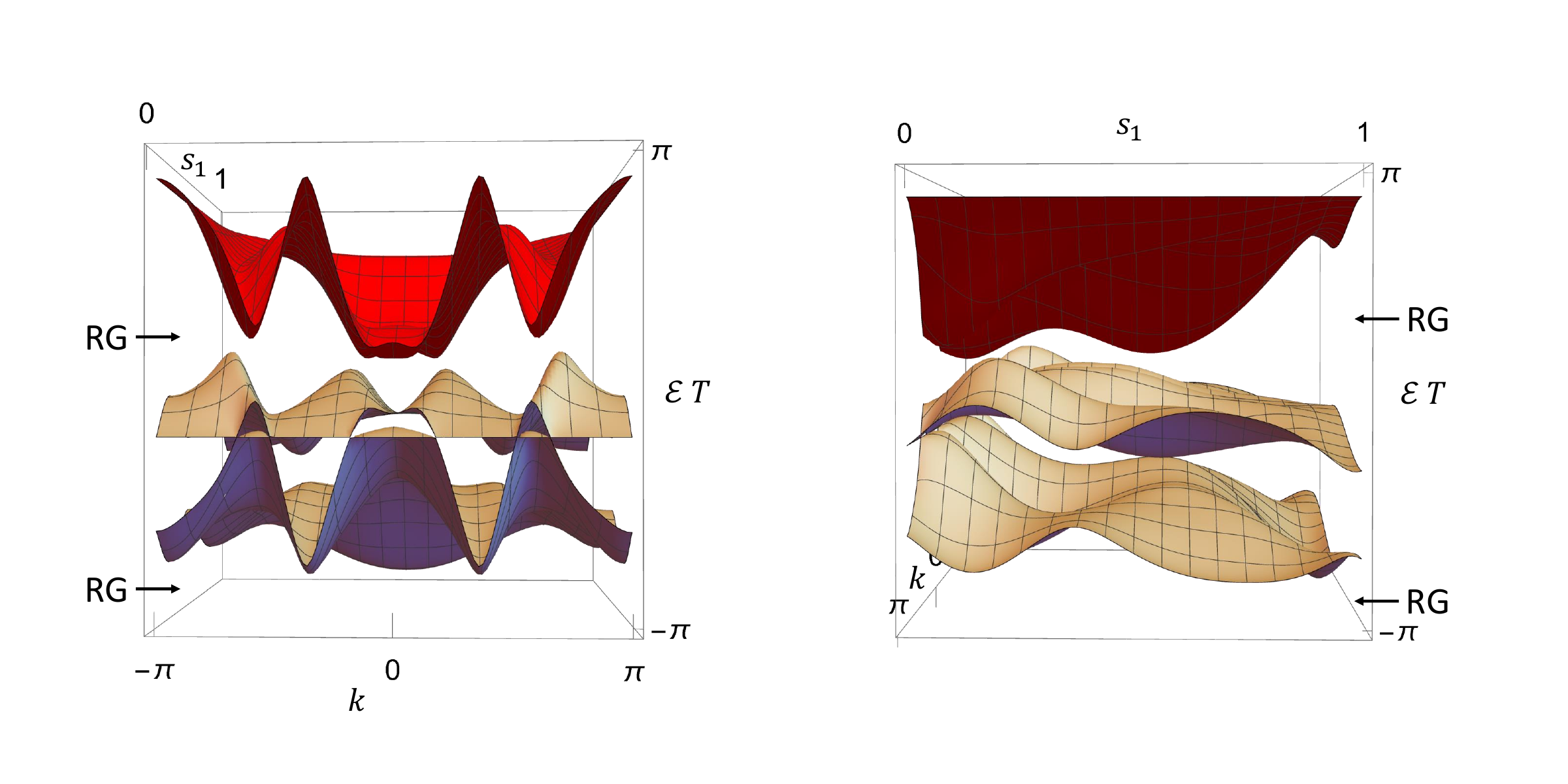}
    \caption{
    The quasienergy bands driven from $\widetilde{U}_{1D,s_1}(T)$ viewed from two angles.
    The deformed PBZ is $[\Phi_{k,s_1},\Phi_{k,s_1}+2\pi )$ with $\Phi_{k,s_1}=-\pi$.
    }
    \label{fig:1D_Inversion_app}
\end{figure}

\end{widetext}

\end{document}